\documentclass[10pt,conference,compsocconf,letterpaper]{IEEEtran}
\usepackage{color}

\usepackage{caption}
\usepackage{subcaption}
\usepackage{cite}
\ifCLASSINFOpdf
   \usepackage[pdftex]{graphicx}
   \graphicspath{{../pdf/}{../jpeg/}}
\else
\fi
\usepackage[cmex10]{amsmath}
\usepackage{times}

\usepackage{color}

\newcommand{\ignore}[1]{}

\newcommand{\squishlist}{
\begin{list}{$\bullet$} {
	\setlength{\itemsep}{0pt}
	\setlength{\parsep}{3pt}
	\setlength{\topsep}{3pt}
	\setlength{\partopsep}{0pt}
	\setlength{\leftmargin}{1.5em}
	\setlength{\labelwidth}{1em}
	\setlength{\labelsep}{0.5em}
	}
}

\newcommand{\squishlisttwo}{
 \begin{list}{$\bullet$}
  { \setlength{\itemsep}{0pt}
     \setlength{\parsep}{0pt}
    \setlength{\topsep}{0pt}
    \setlength{\partopsep}{0pt}
    \setlength{\leftmargin}{2em}
    \setlength{\labelwidth}{1.5em}
    \setlength{\labelsep}{0.5em} } }

\newcommand{\squishend}{
	\end{list}
}

\begin{document}
%\title{Who Should Be Cared in Neighborhood: Social Strength}
%%\title{Exploiting Hidden Social Ties with Social Strength}
%%\title{Quantifying and Exploiting Indirect Social Ties}% with Social Strength}
\title{The Power of Indirect Social Ties}

\author{\IEEEauthorblockN{Xiang Zuo  \hspace{5 mm} Jeremy Blackburn}
\IEEEauthorblockA{Comp. Science \& Eng.\\
University of South Florida\\
Tampa, Florida 33620\\
\{xiangzuo, jhblackb\}@mail.usf.edu}
%%\and
%%\IEEEauthorblockN{Jeremy Blackburn}
%%\IEEEauthorblockA{Computer Science \& Engineering\\
%%University of South Florida\\
%%Tampa, Florida 33620\\
%%Email: jhblackb@mail.usf.edu}
\and
\IEEEauthorblockN{Nicolas Kourtellis}
\IEEEauthorblockA{Yahoo Labs\\
Barcelona, Spain\\
kourtell@yahoo-inc.com}
\and
\IEEEauthorblockN{John Skvoretz}
\IEEEauthorblockA{Sociology Department\\
University of South Florida\\
Tampa, Florida 33620\\
jskvoretz@usf.edu}
\and
\IEEEauthorblockN{Adriana Iamnitchi}
\IEEEauthorblockA{Comp. Science \& Eng.\\
University of South Florida\\
Tampa, Florida 33620\\
anda@cse.usf.edu}
}


%Telephone: (800) 555--1212, Fax: (888) 555--1212}


\maketitle

\begin{abstract}

While direct social ties have been intensely studied in the context of computer-mediated social networks, indirect ties (e.g., friends of friends) have seen little attention.
Yet in real life, we often rely on friends of our friends for recommendations (of good doctors, good schools, or good babysitters), for introduction to a new job opportunity, and for many other occasional needs.
In this work we attempt to 1)~quantify the strength of indirect social ties, 2)~validate it, and 3)~empirically demonstrate its usefulness for distributed applications on two examples. 

%To this end, 
We quantify social strength of indirect ties using a(ny) measure of the strength of the direct ties that connect two people and the intuition provided by the sociology literature. 
We validate the proposed metric experimentally by comparing correlations with other direct social tie evaluators. %\ainote{something specific here}.
We show via data-driven experiments that the proposed metric for social strength can be used successfully for social applications.
Specifically, we show that it alleviates known problems in friend-to-friend storage systems by addressing two previously documented shortcomings: reduced set of storage candidates and data availability correlations. %\ainote{check terminology with papers in the area}.
We also show that it can be used for predicting the effects of a social diffusion with an accuracy of up to 93.5\%. %\ainote{numbers here}.

\end{abstract}


% \fi
\IEEEpeerreviewmaketitle

\section{Introduction}\label{sec:intro}
%Why need n-hop social strength?
%widely utilization of indirect social strength
%direct social ties have been examined by plenty of studies. 
%applications need
%Indirect social ties are proved to be more effective in information diffusion~\cite{}. 

The mining of the huge corpus of social data now available in digital format led to significant advances of our understanding of social relationships~\cite{antonucci1990social} and social behavior~\cite{homans1961social}, and confirmed on larger datasets long standing results from sociology. 
In addition, social information (mainly relating people via declared relationships on online social networks or via computer-mediated interactions) has been successfully used for a variety of applications, from spam filtering~\cite{shen2013spam} to recommendations~\cite{basu1998recommendation} and peer-to-peer backup systems~\cite{li2006f2f}. 

All these efforts, however, focused typically on direct ties.
Direct social ties (that is, who is directly connected to whom in the social graph) are natural to observe and reasonably easy to classify as strong or weak~\cite{kahanda2009transactional, gilbert2009predict}. 
Indirect social ties, though, defined as a relationship between two individuals who have no direct relation but are connected through a third person in their social network~\cite{burt1987social}, carry a significantly larger potential as they facilitate better information dissemination then direct ties~\cite{granovetter1973strength} and enable significantly better opportunities~\cite{granovetter1995job}.
Computer-mediated applications, we conjecture, have a more significant potential in mining and exploiting indirect ties, as the direct ties are likely to be used via the traditional channels through which were used for thousands of years: hopefully people will continue to talk to friends about needs and opportunities without totally relying on online social applications. 

However, not all indirect ties are valuable or usable, even at short distances (i.e., 2 hops). 
For example, a distant acquaintance of a mere acquaintance is unlikely to have the social incentives of doing a personal favor, such as sharing his available storage on his personal computer. 
Moreover, the trust is possibly too diluted in such conditions: why would that weak distant social contact trust that the data he is asked to store is not illegal or malicious? 
In addition, what works for a user or an application might not work for another user or another application: the indirect tie $A$--$B$ may be strong enough for $A$ to use, but not enough for $B$ to use; or it may be strong enough to use for a backup application, but not for a distributed social clustering application. 
Therefore, quantifying the strength of an indirect tie is both necessary and non-trivial. %a challenging problem, especially when the social distance becomes larger (n-hop away, where $n\geq 2$).  

%In addition, provide trustedcan be trusted in the information transmission, some of them have comparable strength as direct ties, others are weakly associated with their social relations. 
%More importantly, the strength of social ties of AB in user A's perspective is different from B's eye---which is more accurate to reflect the real-world relations and trust.
%Little studies reveal the  

In this paper we propose a metric that we call \emph{social strength} that numerically estimates the strength of an \emph{indirect} tie (Section~\ref{sec:metrics}). 
Our metric uses various observations from sociology and builds on the current opportunities of quantifying the strength of direct ties from computer or phone-recorded interactions.
We rely on the sociology literature to define the requirements of such a metric: first, since social relationships are asymmetrically reciprocal~\cite{wellman88analysis}, the social strength of an indirect tie consequently needs to be asymmetrical as well. 
%will likely have different values for the two end points. 
Second, a friend of many of one's friends---thus connected via multiple 2-hop paths---can potentially be more socially ``close'' than the friend of a friend, connected via only one 2-hop path. 
Third, the strength of an indirect tie decreases with the length of the shortest path~\cite{friedkin1983horizons}.

We partially validate the social strength metric of indirect ties  (in Section~\ref{sec:verification}) using real datasets. 
We demonstrate the usefulness of our metric on two proof-of-concept applications: the recruitment of storage candidates from indirect social ties (Section~\ref{sec:f2f}) and the prediction of information dissemination paths (Section~\ref{sec:diffusion}). 
We show experimentally that two main issues identified in friend-to-friend storage systems, namely reduced candidate sets~\cite{li2006f2f} and low availability due to time synchronization among friends~\cite{raul2012peer}, are significantly alleviated by employing our social strength metric for the recruitment of socially close indirect contacts.
Specifically, online availability of storage resources are improved by up to 20\%.
%\ainote{numbers here from F2F}
We also show  that the  social strength metric predicts with an accuracy of up to 93.5\% diffusion paths 2-3 steps ahead to provide more time for decision makers for containing damaging information dissemination (i.e., damaging rumors) or accelerating information spreading. 

\section{Related Work}

Since Granovetter~\cite{granovetter1973strength} introduced the notion of strength of ties in social networks, there have been many studies on tie strength measurement.
%A number of tie strength measures were discussed in pervious literature.
Common Neighbors~\cite{gupte2012implicit} is a measure of tie strength that considers the number of common activities or friends that two users share.
Jaccard index~\cite{jaccard1901distribution} was shown to estimate the number of phone calls between two users by considering their shared neighborhoods~\cite{onnela2007structure}. 
%\jbnote{Would be nice to find a better citation. This one is pretty old and not related to social networks from what I can tell.} is a more refined measure of tie strength, given by the overlap divided by the union of two users' social circles.
%hile simple, without considering more information from users, these metrics are not sufficient to precisely estimate the strength of people's social ties. 
%\jbnote{More complex applications than what?}

Homophily, or similarity between individuals, has been shown to be a catalyst for the formation of relationships~\cite{mcpherson2001birds} and has been leveraged to predict the existence and strength of social ties.
% It has been leveraged users' homophily (similarity) to predict the existence of social ties. 
Gilbert and Karahalios~\cite{gilbert2009predict} modeled tie strength as a combination of social dimensions such as intensity, intimacy, duration, and structure.
Crandall et al.~\cite{crandall2010inferring} investigated the existence of social ties between people from co-occurrence in time and space on Flickr
%The authors used geotags and timestamps to infer whether two users were at the same location within a certain time period.
and discovered that even a small number of co-occurrences indicate a high probability of an existing tie between two users.
Likewise, 
%Wang et al.~\cite{} proposed using an unsupervised algorithm to infer advisor-advise relationships from a publication network. 
%Tang et al.~\cite{tang2012heterogenous} developed a framework for inferring implicit social ties by learning across heterogeneous network. 
%However, only predict the existence of indirect social ties is not enough for current applications. 
%Therefore,
Kahanda and Neville~\cite{kahanda2009transactional} developed a supervised learning predictor that classifies a link in OSNs as either a weak or strong tie via features from user profiles, graph topology, transactional connectivity and network-transactional
connectivity features.
%finding that network-transactional connectivity was the most important predictor.\jbnote{What is network transactional connectivity? Does it relate to the current work?}
Adamic and Adar~\cite{adamic2003friends}  introduced a log-based similarity metric to capture tie strength between indirectly connected nodes.

However, these methods either need extra information---such as users' profiles, the message content or users' geo-locations---or adopt complex models that cannot be implemented in a decentralized fashion.
More importantly, most previous methodologies simply treat users' relationships symmetrically.
Without asymmetric discrimination, it is difficult to accurately capture the strength of social ties~\cite{brass1998relationships}.

This work builds on our preliminary definitions~\cite{anderson2010onmanaging,kourtellis2012onthedesign} of social strength metric but significantly changes them by introducing new lessons from sociology.
Most importantly, this work contributes the validation of the social metric and demonstrates its value via proof-of-concepts applications that use it.

%=================================
\section{Social Strength Definition}\label{sec:soc-strength}
%=================================
\label{sec:metrics}
%\newpage
%\clearpage

%\subsection{Quantifying Social Strength}

We want to define a metric that quantifies the intensity of a social connection between indirectly connected nodes in a social network. 
The need for such a metric is intuitively supported by many sociological studies and is also well understood from daily life: friends of friends are an important resource for  information and useful social contacts.

In our attempt to quantify an indirect social tie, we use the following observations from sociology and from recent data-driven studies on computer-mediated social relationships:

\squishlist
\item[O1:] The strength of a direct social relationship is influenced by the amount of interactions, as confirmed in~\cite{granovetter1973strength,xiang10modeling}. 
Moreover, interactions among OSN users were shown to represent more meaningful relations than just declared relationships~\cite{wilson2009userinteractions}.
Consequently, in the quantification of an indirect social tie, we rely on a numerical representation of the strength of a \emph{direct} social tie, that can be expressed as number of interactions, number of shared interests, or other recordable outcomes, depending on the semantic of the relationship. 

\item[O2:] 
Intuitively, the social strength of an indirect relationship on a particular social path is limited by the strength of the weakest direct tie on that path. 
%Also intuitively, 
Furthermore, the strength of an indirect tie decreases with the length of the shortest path between the two individuals. 
This has been qualitatively observed by Friedkin~\cite{friedkin1983horizons}, who concluded that the horizon of observability is limited to a distance of 2 hops.
Without contradicting this result, in this work we propose that instead of limiting indirect social ties to a distance of 2, we allow the calculation of social strength along longer distances, but decrease its value as a function of distance. 

\item[O3:] 
Multiple types of social interactions (for example, both professional collaboration and playing tennis after work) result into a stronger (direct) relationship than only one type of interaction~\cite{pappalardo2012multidimensional}.
Furthermore, sociology studies~\cite{friedkin1983horizons} observed that %two individuals have meaningful social relationships when connected within two social hops (i.e., ``horizon of observability''), 
%\xznote{shall we need to change a way of saying?} and 
the relationship strength of indirectly connected individuals greatly depends on the number of different direct or indirect paths connecting them.
Therefore, we consider the strength of multiple shortest paths in our definition of the strength of an indirect social tie.

\item[O4:] 
Typically, social ties between individuals are asymmetrically reciprocal~\cite{wellman88analysis}.
%For example, employees trust their boss more than the boss trusts his or her employees.
%This is seen in a variety of situations.
%Because individuals have d
Thus, for the directly connected users Alice and Bob, the importance of their mutual relationship may be dramatically different. 
Different experiences, psychological backgrounds and personal histories cause these asymmetries~\cite{golbeck2006inferring}.
%Different experiences, roles and the power dynamics in the relationship cause these asymmetries~\cite{}.
We want to preserve this asymmetry in quantifying indirect ties, such that Alice and Charlie, indirectly connected via Bob, are entitled to have different views about their indirect tie. 

%%Some socially aware applications might be able to make use of longer social distances, while others will restrict their domain to closer indirect ties. 

\squishend

Therefore, to quantify the social strength of an indirect social tie between users $i$ and $m$ we consider relationships at any $n$ ($n\geq 2$) social hops, where $n$ is the shortest path between $i$ and $m$. 
We assume a weighted interaction graph model that connects users with edges weighted based on the intensity of their direct social interactions.
%To this end, we adapt the \textit{Social Strength(A,B)} function~\cite{anderson2010onmanaging, kourtellis2012onthedesign} that returns a real number between 0 and 1 that quantifies the social strength between users $A$ and $B$ from $A$'s perspective.
Assuming that $\mathcal{P}_{i, m}^n$ is the set of different shortest paths of length $n$ joining two indirectly connected users $i$ and $m$ and $\mathcal{N}(p)$ is the set of nodes on the shortest path $p, p \in \mathcal{P}_{i, m}^n$, we define the social strength between $i$ and $m$ from $i$'s perspective over an $n$-hop shortest path as:
%%\begin{figure*}
\begin{equation}
%\begin{split}
SS_{n}(i,m) = 1- \prod_{p \in \mathcal{P}_{i,m}^n} (1- \frac{\min\limits_{j,...,k\in \mathcal{N}(p)}{[NW(i,j),...,NW(k,m)]}}{n})
\label{eq:socs}
%\end{split}
\end{equation}
%%\end{figure*}

This definition uses the normalized direct social weight $NW(i,j)$ between two directly connected users $i$ and $j$, defined as follows:

\begin{equation}
%NW(i, j) = \frac{\sum\omega(i, j)}{\max_{\forall j \in J_i} (\sum \omega(i,j))}
NW(i, j) = \frac{\sum_{\forall \lambda \in \Lambda_{i,j}}\omega(i, j, \lambda )}{\sum_{\forall k \in N_i}\sum_{\forall \lambda \in \Lambda_{i,k}} \omega(i,k,\lambda)}
\label{eq:nw}
\end{equation}

Equation~\ref{eq:nw} calculates the strength of a direct relationship by considering all types of interactions $\lambda \in \Lambda$ between the users $i$ and $j$ such as, let's say, phone calls, interactions in online games, and similar ratings on Netflix (observation O3). 
These interactions are normalized to the total amount of interactions of type $\lambda$ that $i$ has with other individuals.
This approach ensures the asymmetry of social weight (observation O4) in two ways: first, it captures the cases where $\omega(i, j, \lambda) \neq \omega(j, i, \lambda)$ (such as in a phone call graph). 
Second, by normalizing to the number of interactions within one's own social circle, the relative weight of the mutual tie will be different for the two users involved even in undirected social graphs (that is, when  $\omega(i, j, \lambda) = \omega(j, i, \lambda)$). 

The implementation of observations O1, O3 and O4 in the definition of the $NW$ function is naturally carried over in the definition of social strength from Eq.~\ref{eq:socs}.
Moreover, O3 is additionally implemented by considering the product over all shortest paths $p$ that connect two users. 
O2 is implemented by considering the weakest link (minimum normalized weight of all direct ties on each path) and by dividing it with the distance $n$ between the users. 

\ignore{ 
We normalize all the edge weights to help us scale the social strength among different users.
Also, this normalization results in asymmetrical relationships between user i and j. 
The inherent reason is that the normalization is relevant to each user's degree and edge weight. 

For denominator, $N_i$ is the set of directly connected neighbors from $i$ to others. 
Denominator sums all edge weights from user i to all other connected users with all labels.

In our interaction graph model, the weighted edges enable us to capture this asymmetrical behavior.
Since the social strength function normalizes a user's edge weights with his local edge weights aggregation, the asymmetric attribute of relationships is maintained.
This calculation is local to each user, and thus can be implemented fully decentralized.
}

%\subsection{Advantages of Social Strength Metrics}

The proposed social strength measure can:

\squishlist
%\item assign a value for indirect tie strength for nodes indirectly connected at any hop distance
\item Quantify the indirect tie strength for nodes indirectly connected at any social distance. %instead of only a strong or weak tie.
%\item \ainote{rephrase? hard to parse. In fact, I don't understand}	discriminate among the nodes to which a focal node is indirectly connected through the different values of indirect tie strength to those indirect contacts
\item Treat indirect ties between two nodes as possibly asymmetric in strength rather than constraining the values to be equal.
%\item \ainote{what are we saying? how else? do we refer to other situations in the literature?} be calculated using only the information available in the social graph
\item Be more sensitive to strength differences because it uses both edge weights and number of paths to calculate a value.
\item Be calculated locally and so implemented on very large graphs in parallel easily.
\squishend

\ignore{
	\item social strength gives an asymmetric social relationships instead of equal strength for a pair of users.
	\item no extra information outside social graphs are needed in calculation social strength values.
	\item by considering both edge weights and the number of paths between two users, social strength defines social ties more accurate.
	\item the entire calculation is local to each node and thus can be implemented parallel.
	\item can be widely used in various applications.
	
\end{itemize}
}

\section{Social Strength Verification}\label{sec:verification}

Studies show that interviews meant to quantify the strength or even the existence of social relationships are unreliable~\cite{Killworth1990289}, as subjects do not accurately recall~\cite{russell1984problem} or are unable to objectively asses their social relationships.
To verify the social strength metric proposed in Secion~\ref{sec:soc-strength}, we thus chose to quantitatively compare social strength with other metrics, even when the comparison can be done on a limited domain. 
We identified two such metrics of relevance: the overlap of the social neighborhoods and the frequency of interactions in online social networks. 
However, both have limited applicability:
The overlap of social neighborhoods is possibly non-empty only for nodes at distance at most 2. 
The frequency of interaction is by definition only possible for direct social interactions, thus directly connected nodes.

The strength of a relationship between two nodes in a social graph was shown to correlate with the overlap of their social neighborhoods in various studies.
Intuitively, the more friends in common, the closer the relationship between two subjects. 
This hypothesis was verified at scale by studying a who-talks-to-whom real-world mobile phone network~\cite{onnela2007structure}, and demonstrating that neighborhood overlap increases with increasing tie strength.
%\ainote{more specifics here and references }
%%%\xznote{The study in sociology~\cite{granovetter1995job} states that the strength of a tie between two users increases with the overlap of their friend circles. In literature of computer science, people also verified this hypothesis by using a real-world mobile phone network~\cite{onnela2007structure}, and even demonstrated neighborhood overlap of edges is a almost linear function of their tie strength~\cite{easley2010networks}.}

Interaction graphs were shown to provide more accurate representation of social ties than just the existence of declared relationships in OSNs~\cite{huberman2009twitter, wilson2009userinteractions}.
%\xznote{I do not know the second reference.}
%\ainote{do we have more to say about this? Fill a short paragraph}
As social interactions always require some kind of investment of time and effort from participants, the frequency of interaction is an informative measure of tie strength.
Gilbert et al.~\cite{gilbert2009predict} classified tie strengths according to users' attributes on social media such as intimacy and interaction intensity. 
Likewise, Marlow et at.~\cite{economist2009primates} investigated users' tweet and retweet interactions on Twitter to distinguish a user's strong ties from weak ties over an observation period.
% \ainote{specifics?}.

Thus, both the neighborhoods overlap and the frequency of interactions between two users are considered quantifiers of social ties. % among people.
In this section, we compare our proposed social strength metric with these two accepted metrics. % on three real-world social networks. 

%(an indirect social tie indicator) on real-world online social networks by comparing the correlations to neighborhood overlaps and the frequency of interactions with Pearson Correlation Coefficient (PCC)~\cite{rodgers1988correlation}. 

%\vspace{-7mm}

%===========================================================
\subsection{Datasets}\label{sec:datasets}
%===========================================================

We used three social networks for the validation of the social strength metric and for the rest of the experimental evaluations in this paper. 
The first two (CA-I and CA-II) are co-authorship networks from ArnetMiner~\cite{arnetminer} and the third (TF2) is derived from gameplay logs from a Team Fortress 2 online gaming server. 

%\vspace{-2mm} \ainote{don't fuss with the format}

%\subsubsection{Social Networks} \ainote{what's the point of this subsection?}

ArnetMiner mines the academic social network to provide domain-specific search services for researchers.
From this service, Tang et. al~\cite{tang2009social} extracted a weighted co-authorship graph of Computer Science researchers from a variety of domains.
Nodes in this graph represent authors and are labeled with the author's affiliation.
Edges exist if the two researchers co-authored at least one paper together and are weighted with the number of papers co-authored.
The dataset also provides the authors' affiliations. %, from which we extract timezone information.
From this dataset we extracted two networks: \emph{Co-authorship I (CA-I)} is a small connected component with 348 nodes and a relatively low density (see Table~\ref{table:network-characteristics}).  
\emph{Co-authorship II (CA-II)} is the largest connected component of the ArnetMiner co-authorship network, comprising 1,127 nodes and having a density one order of magnitude higher than CA-I.

Team Fortress 2 is an objective-oriented first person shooter game released in 2007. 
We obtained just over 10 months of gameplay traces (from April 1, 2011 to February 3, 2012) from a Team Fortress 2 server located in Los Angeles, California~\cite{blackburn2013relationships}. % , which is owned and operated by the ``Brotherhood of Slaughter'' gaming community.
The logs include game-based interactions among players, such as teammates capturing territory together, or players from opposing teams ``killing'' each other. % on the same team working together to ``kill'' an opponent. 
In addition, we crawled the \emph{Steam Community} online social network~\cite{steam} and obtained the set of friends for each player in the server logs~\cite{blackburn2012scarletc}.
We were able to tie a player on the server to an OSN profile because Team Fortress 2 uses the Steam gaming platform, which includes Steam Community, among other gaming specific services.
The \emph{Team Fortress 2 interacting friends network (TF2)} is composed of edges between players who had at least one in-game interaction while playing together on this particular server, and also have a declared friendship in Steam Community. %\ainote{one could ask why the choice of this network from the possible ones: }
This dataset has two advantages over the Steam declared OSN: First, it provides the number of in-game interactions that quantifies the strength of a declared social tie.   
Second, it provides players' online/offline status that we use later in the experiments in Section~\ref{sec:f2f}.
Over a pure in-game interaction network, it provides the advantage of selecting the most representative social ties, as proven in~\cite{blackburn2013relationships}. 
In this network of 2,406 nodes and over 9,000 edges, edge weights thus represent the number of in-game interactions.

\begin{table*} 
\begin{center}
%\itshape
\caption{Characteristics of the social networks used in experiments.}
\label{table:network-characteristics}
\begin{tabular}{|l|r|r|r|r|r|r|r|r|}
\hline
Networks & \# Nodes & \# Edges & Avg. Path Length &Density & Clustering Coef. & Assort. & Diam. & Range edge weights\\
 \hline\hline 
 CA-I & 348& 595 & 6.1 &  0.0098 & 0.28 & 0.173 & 14 &[1--52]\\ 
 \hline
 CA-II &1,127 & 6,690 & 3.4&  0.0100 & 0.33 & 0.211 & 11& [1--127] \\ 
\hline 
TF2 &  2,406 & 9,720 & 4.2 & 0.0034& 0.21 & 0.028 & 12 & [1--21,767]\\
\hline

\end{tabular}  
\end{center}

\end{table*}

\begin{figure*}[tb]
\begin{center}
      \graphicspath{{./graph/}}
      \includegraphics[scale=0.35]{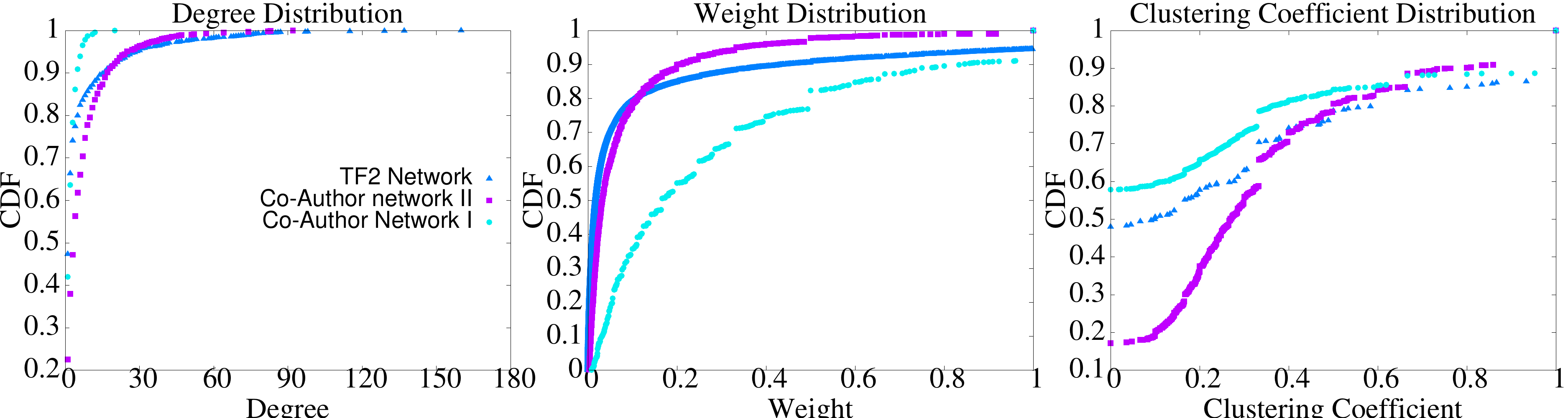}
      \caption{\label{all-distribution} CDF of the three networks' degree, weight and clustering coefficient distributions.
      Edge weight distributions are normalized to allow for comparison among the three networks.}
\end{center}
\end{figure*}
%\vspace{-2mm}

A brief characterization of the networks appears in Table~\ref{table:network-characteristics}.
Figure~\ref{all-distribution} plots the degree, edge weight, and clustering coefficient distributions for each of our networks.
In order to compare the weight distribution between different networks, we normalized all edge weights by dividing them to the largest edge weight in the corresponding network. % between 0 and 1. 

%===========================================================
\subsection{Social Strength vs. Neighborhood Overlap}
\label{sec:ss-vs-neighborhood}
%Correlations Between Neighborhood Overlap and Social Strength}
%===========================================================

The overlap between the social neighborhoods of two users $s$ and $r$ can be represented by the Jaccard coefficient defined as follows: %which considers the overlap of two users' neighbors over the union of their neighbors.
%The definition of JC is as follows:

\begin{equation}
JC(s, r) = \frac{n_{sr} }{degree(s) + degree(r) - n_{sr}}
\end{equation}
where $n_{sr}$ is the number of mutual neighbors of nodes $s$ and $r$, and $degree(s)$ and $degree(r)$ are the number of edges of nodes $s$ and $r$, respectively.

There are some immediate observations:
 $JC(s, r)$ is symmetrical, that is,  $JC(s,r) = JC(r,s) $. %to a pair of users. I.e., the JC is the same from user i to j and user j to i.
Also, $JC(s,r)=0$ for all nodes $s$ and $r$ situated in the network at a distance larger than 2 hops. %metric is limited to at most with 2-hop relationships, so beyond 2-hop distance, the JC value is zero.
Meanwhile, social strength is used for quantifying the strength of indirect social ties of at least 2 hops distance in the social graph.
Therefore, for a meaningful comparison between $JC(s,r)$ and the social strength between the same nodes $s$ and $r$, $SS_n(s,r)$, we select only those nodes $s$ and $r$ for which $n=2$. %that are 2 hops away. %quantify both values on 2-hop social distance that is depicted in Figure~\ref{corr1}.
%\begin{figure}
%\begin{center}
  %    \graphicspath{{./figure/}}
    %  \includegraphics[scale=0.4]{corr1.pdf}
      %\caption{\label{corr1} 2-hop length path between A and B.}
%\end{center}
%\vspace{-2mm}
%\end{figure}

Table~\ref{tab:pearson-stats-degreejaccard} shows the Pearson correlation coefficient between $JC(s,r)$ and $S_2(s,r)$ for all pairs of nodes $(s,r)$ situated at distance 2 in the social graphs (all p-values $\leq2.2\times10^{-16}$).
All three datasets present positive correlations. 
In the CA-II network the correlation is the highest ($0.41$), while TF2 shows only a weak positive correlation.
Because TF2 is extracted from an online gaming social network, players establish connections according to their gaming interests or to the requirements of the game, rather than according to genuine social phenomena.
%\xznote{
For example, many players add others as their social connections to satisfy the minimum number of participants required for playing the game, but end up having few interactions. 
%In fact, for such relationships the social overlap is very low, so Jaccard coefficient cannot precisely capture the strength of social ties.}
Thus, in this scenario, the overlap of neighborhoods may not accurately capture meaningful social ties without considering the frequency of users interactions. %\ainote{I cannot follow the argument. It seems to me that, on one hand, we could verify if the Jaccard coefficient of the TF2 network is different than in genuinely social networks (I don't know how -- maybe looking at distributions?. On the other hand, I don't understand the leap from Jaccard coefficient to frequency of social interactions -- I think you're saying that the network structure may be social with direct ties but not so social 2-hops away? We need a proof for this argument, if we are to advance it.}

In contrast, social strength evaluates social closeness by including both shared paths and interactions between users, which calibrates social ties more meaningfully.
 
\begin{table} 
\begin{center}
\caption{Pearson correlation coefficients (PC) between the Jaccard coefficient and social strength over 2-hop paths.}\label{tab:pearson-stats-degreejaccard}
\begin{tabular}{|c|r|}
\hline
Networks &PC(JC, $SS_2$) \\
 \hline\hline 
 CA-I & 0.255\\ 
 \hline
 CA-II & 0.410  \\ 
\hline 
TF2 & 0.137  \\ 
\hline 
\end{tabular}  
\end{center}
 %\vspace{-6mm}
 \end{table}

%=========================================================================================
\subsection{Social Strength vs. Number of Interactions}%Correlations among Interactions and Social Strengths}
\label{sec:1-hop-verification}
%=========================================================================================

\begin{figure}
        %\centering
         \graphicspath{{./figure/}}
        \begin{subfigure}[]{0.10\textwidth}
                \includegraphics[width=\textwidth]{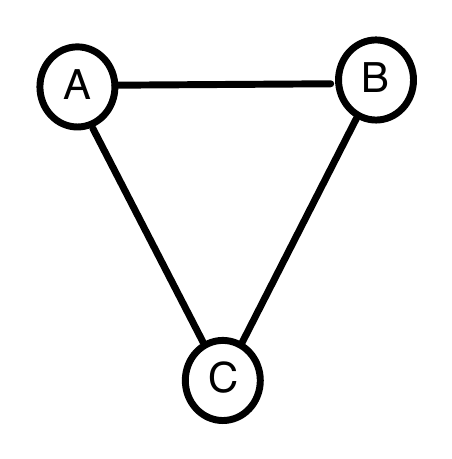}
                \caption{}
                \label{a}
        \end{subfigure}
        ~%add desired spacing between images, e. g. ~, \quad, \qquad etc.
        \begin{subfigure}[]{0.10\textwidth}
                \includegraphics[width=\textwidth]{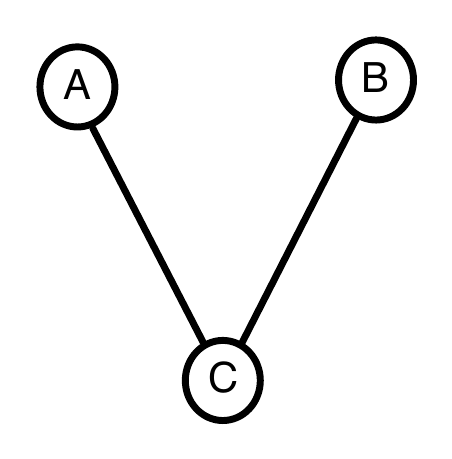}
                \caption{}
                \label{b}
        \end{subfigure}
       ~%add desired spacing between images, e. g. ~, \quad, \qquad etc.
        \begin{subfigure}[]{0.10\textwidth}
                \includegraphics[width=\textwidth]{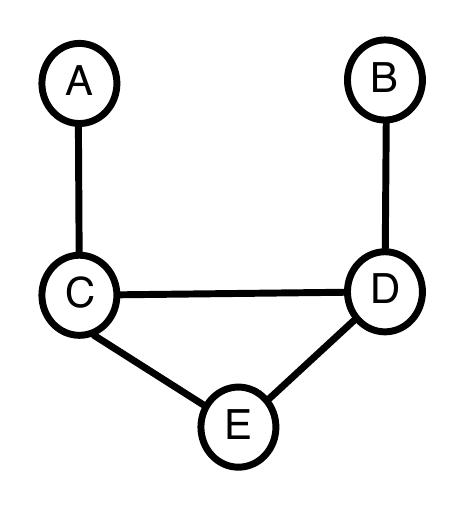}
                \caption{}                
                \label{c}
        \end{subfigure}
          ~ %add desired spacing between images, e. g. ~, \quad, \qquad etc.
        \begin{subfigure}[]{0.10\textwidth}
                \includegraphics[width=\textwidth]{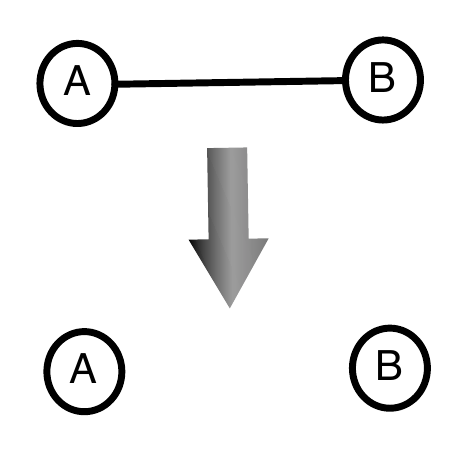}
                \caption{}
                \label{d}
        \end{subfigure}
        \caption{Using 2-hop social strength metrics to infer 1-hop social strength values: (a) original graph (b) remaining graph after removing edge AB (c) graph without 2-hop length paths (d) graph with only one shared edge between A and B.}\label{fig:ss-cal}
\end{figure}

A common measure of the strength of a direct social tie is the number/frequency/duration of interactions~\cite{granovetter1973strength, gonalves2011dunbar}. 
In order to compare social strength (a measure of the strength of an indirect social tie) with the strength of a direct tie as given by the number of direct social interactions, we do the following: for closed triads in 
the social graph (such as depicted in Figure~\ref{a}), we calculate the correlation between the social strength of $A$ and $B$ along the path $A-C-B$ by ignoring the direct tie $A-B$ (as depicted in Figure~\ref{b}) and the number of interactions between $A$ and $B$. 
Because the overlap of the social neighborhoods of directly connected nodes $A$ and $B$ is also a measure of the intensity of the social tie~\cite{onnela2007structure}, we also include the Jaccard coefficient of the direct tie $A-B$ in our comparison. 
(Note that this time we apply the Jaccard coefficient to direct ties, not to 2-hop distant nodes, as in the previous section).

%In social graphs, edge weights could be the frequency of users' interactions or other meaningful numbers to infer the closeness relationships among users.
%In our datasets, all the edge weights are the number of frequency among users, either number of co-authored academic paper or variety of game-related interactions in the online game network.
%Instead, we examine the correlation among edge weights and \emph{Social Strength} values. 
%To present a comprehensive comparison, we also include neighborhood overlap in the comparison.

%However, all these three values are not with the same path lengths: edge weights are all 1-hop paths, jaccard coefficient can be either 1- or 2-hop path lengths and social strengths apply to at least 2-hop paths.
%In order to compare all these values in the same condition (1-hop path), we use 2-hop social strengths to indicate 1-hop social strength values. 
%To do this, we remove the directly connectedly edge, if it exists, between a pair of users A and B as shown in~\ref{a} of Figure~\ref{fig:ss-cal}, then calculate 2-hop \emph{Social Strength}(A, B) with the remaining edges in~\ref{b}. 

The intuition for this experiment is the following:  if the social strength metric indeed captures the strength of an indirect tie between $A$ and $B$, then this metric should correlate with measures of strength on the direct tie $A-B$. 
%is that \ainote{Hmm... what's the intuition?} 
In Table~\ref{tab:correlation}, column 2 to 4 present the Pearson correlation among the number of direct interactions, 1-hop Jaccard coefficient, and the social strength value along n-hop (n is the shortest path length)
%2-hop 
paths for pairs of directly connected users.
The results show that all correlations are positive (with p-values $\leq 8.8 \times 10^{-8}$).
Moreover, it results that social strength is a better predictor of the intensity of a social relationship than the Jaccard index: social strength has higher positive correlation with edge weights than the Jaccard coefficient in all three datasets, with the highest positive correlation value in the CA-II network.
This result is mainly due to the fact that the Jaccard coefficient metric does not takes edge weights into consideration, while social strength considers both edge weights and shared paths for quantifying users' social closeness.

%\xznote{
However, removing edges such as $A$--$B$ could lead to situations such as those in Figures~\ref{c} and~\ref{d}, where $A$ and $B$ get disconnected or remain connected by shortest path of 3 or more. %but connect by 3- and 4-hop paths or no link between them. 
In the case of Figure~\ref{c}, the Jaccard coefficient becomes $0$ while the social strength can still be computed as $SS_3(A,B)$ along the path $A$--$C$--$D$--$B$. % by 3- or 4-hop social distance. 
%As social ties decrease with the increase of social distance, we choose calculate the social strength values via the shortest length path, which is the strongest social tie indicator, e.g., \emph{$Social Strength(A,B)$} = {$Social Strength_3(A,B)$}, which is computed through the path $A-C-D-B$ in ~\ref{c}.
%}
In the case of Figure~\ref{d}, both social strength and the Jaccard coefficient become $0$, which might bias the results. 
%\ainote{but it's not true, since ss can be theoretically computed along a n-hop path. Why do we limit ourselves to computing 2-hop ss?}
%\xznote{Yes, SS can be calculated by 3-, 4- or 5-hop path. But things will be very complex as SS could have all 3-,4- and 5-hop values, then how to decide the final value of SS? simply plus all these three values, multiply them or select the maximum one? At least, I think in most cases, 2-hop is the strongest value among all available values. The exception is that large edge weights and many shared paths between two nodes can generate high social strength value for more than 2 hops. But if this is only valid for a small portion of node pair in the graph, the correlation should be very low.}
%Also, the comparable high correlation between the JC and \emph{Social Strengths} in the fourth column is mainly caused by the special scenarios in~\ref{c} and \ref{d} of Figure~\ref{fig:ss-cal}, where when \emph{Social Strengths} values decrease to zero the JC values also go to zero.
Column 5 to 7 in Table~\ref{tab:correlation} present the correlations among the three metrics after removing all such zero values in each dataset (all p-values $\leq 0.01$). 
The percentage of removed data is presented in the 8th column. 
Even in this case when we remove all the special zero values and re-calculate the Pearson coefficient, social strength values rarely change and have higher correlation with edge weights than the Jaccard coefficient has. 

% then both the JC and social strength values are zeros, and if the only shared path (bridge) between A and B is removed the \emph{Social Strength} value is also zero.
%%We eliminate such cases from our calculations, because we cannot compute social strength in situations such as Figure~\ref{d} or we artificially make the Jaccard coefficient $0$, as in Figure~\ref{c}. 
%%(Because of these eliminations, reported the correlation numbers

%\begin{table} 
%\begin{center}
%\itshape  
%\caption{PCC among edge weights, Jaccard coefficient and the Social Strengths in 1-hop paths with all cases.}\label{tab:correlation2}
%\begin{tabular}{|l||c|c|c|}
%\hline
%Networks &Cor(weight, jc) & Cor(weight, ss) & Cor(jc, ss)\\
 %\hline\hline 
 %CA-I & 0.179& 0.238 & 0.655 \\ 
 %\hline
 %CA-II &0.211  & 0.394 & 0.591\\ 
%\hline 
%TF2 & 0.173  &  0.267 & 0.554\\ 
%\hline 
%\end{tabular}  
%\end{center}

 %\end{table}
  
%\begin{table*} 
%\begin{center}
%\itshape  
%\caption{PCC among edge weights, Jaccard coefficient and the Social Strengths in 1-hop paths without special cases.}\label{tab:correlation3}
%\begin{tabular}{|l||c|c|c|r|}
%\hline
%Networks &Cor(weight, jc) & Cor(weight, ss) & Cor(jc, ss) & Per. of Removed Zeros\\
 %\hline\hline 
 %CA-I & 0.173& 0.244& 0.530 & 24.7\% \\ 
 %\hline
 %CA-II & 0.210  & 0.396 & 0.580 & 2\% \\ 
%\hline 
%TF2 & 0.135  &  0.245 & 0.492 & 13.2\%\\ 
%\hline 
%\end{tabular}  
%\end{center}

 %\end{table*}

%%%%%%%%%%%%%%%%%%%%%%%%%%%%%%%%%  merged tables
\begin{table*} 
\begin{center}
%\itshape  
\caption{Pearson coefficient (PC) among edge weights, Jaccard coefficient and the social strength over 1-hop paths and the same correlations by removing zeros in special cases.}\label{tab:correlation}
\begin{tabular}{|l||c|c|c|c|c|c|c|}
\hline
Networks &PC(weight, JC) & PC(weight, $SS_1$) & PC(JC, $SS_1$) & PC\_nonzero(weight, JC) & PC\_nonzero(weight, $SS_1$) & PC\_nonzero(JC, $SS_1$) & \% of Removed Zeros\\
 \hline\hline 
 CA-I & 0.179& 0.238 & 0.655 & 0.173& 0.244& 0.530 & 24.7\%  \\ 
 \hline
 CA-II &0.211  & 0.394 & 0.591 & 0.210  & 0.396 & 0.580 & 2\%  \\ 
\hline 
TF2 & 0.173  &  0.267 & 0.554 & 0.135  &  0.245 & 0.492 & 13.2\% \\ 
\hline 
\end{tabular}  
\end{center}

 \end{table*} 
 
To conclude, the above experiments show that our definition of social strength $SS_n$ among indirectly connected individuals in a social graph is positively correlated to both neighborhood overlaps and the frequency of interactions.
%This tells us when the neighborhood overlaps or the frequency of interactions are high, social strength is also high.
Consequently, since the neighborhood overlap and the frequency of interactions are considered to accurately estimate the strength of social ties, then $SS_n$ can also be used for indicating the closeness of social relationships among people.
The difference is that the neighborhood overlap and the frequency of interactions are limited to quantifying the intensity of direct or at most 2-hop away social ties, while the social strength can estimate the intensity of ties between people who are n-hop distant in the social graph. 
%\ainote{Hmm... this doesn't really fly, because we only showed that 2-hop ss works somewhat. Not 3-hop ss or longer.}
%\ainote{there is also another difference: the ss is asymmetrical. Did we consider this in the experiments? That is, did we calculate both ss(A-B) and ss(B-A)? If not, how did we decide for which direction to consider? Is there something we can do about this? }
%\xznote{Yes. I considered the asymmetrical social strength values when calculating correlations. For Jaccard and edge weights, I copied the value of AB to BA (as they are symmetrical).}
%\xznote{I did an experiment tonight to compare the correlation between edge weights and 3-hop social strength on CA-I dataset. The Pearson coefficient is 0.05622 with p-value = 0.7679. It shows very weak positive correlation and high p-value that both are not good results. The reason is 3-hop social distance is much weaker than 2-hop to evaluate the strength of social ties in CA-I dataset. I also checked the 3-hop social strength values, and the largest one is 0.047, others are all very small numbers. It might because 3-hop social strength values are too small even if the corresponding edge weights are high, the social strength values have not much changed. These 3-hop social strength values are calculated based on removing the directly connected edges.}

%\section{The Power of Social Strengths in Socially Aware Applications}
%give examples about the widely usage of social strength.
%====================================================
\section{Using Social Strength in Friend-to-Friend Storage Systems}
%The Power of Social Strength in Improving Service Performance of Friend-to-Friend Storage Systems}

\label{sec:f2f}
%====================================================

A Friend-to-Friend (F2F) storage system is a distributed system where users use incentives to get access to the available storage resources or services of their friends' machines. 
While a promising alternative to cloud-based data backup, F2F storage systems were shown to suffer from two significant limitations.
First, users with a small set of friends are penalized by lack of available storage for their needs~\cite{li2006f2f}.
Second, friends are typically in close geographical proximity, and thus their online times are synchronized, leading to high unavailability to their friends' data~\cite{raul2012peer}. 
These concerns can intuitively be addressed by leveraging social strength $SS_n$ (where $n \geq 2$) to expand the set of storage resources while still using a measure of social incentives. 

In this section we verify whether $SS_n$ can improve the service performance in F2F storage systems.
To maintain a meaningful value of social incentives, we restrict our evaluations to $n=2$ and $n=3$. 
%\subsection{Experiment Design}
%We mainly focus on quantifying the size of the expanded friend sets, and analyzing the system performance in terms of potential data availability on the three datasets, CA-I, CA-II and TF2.
Our objectives are:
%\begin{enumerate}
\squishlist
	\item To understand if $SS_n$ expands the size of candidate sets.
	\item To evaluate the benefits of using $SS_n$ to improve data availability in F2F systems. % when taking empirical online presence patterns into account.
\squishend
%\end{enumerate}

%-----------------------------
\subsection{Experimental Setup}
%-----------------------------

Details of algorithmic and empirical assumptions made in our experiments are presented in the following. 

\subsubsection{Expansion algorithm}
\label{sec:expansion-alg}

A user is not expected to trust all his friends of friends to store his or her data: some of them can be just weak connections of his weaker direct social ties, for example. %weakly connected to his 1-hop friends, and some others may have very few common friends with him. 
We use the quantitative power of $SS$ to select the reasonably strong indirect connections and guarantee comparable social strength with the user's 1-hop friends. 

The expansion algorithm follows two steps:
%\begin{enumerate}
\squishlist
\item For each user $i$, find the weakest direct social contact $p$ such that $NW(i, p) = \min\limits_{j \in Neigh(i)}{[NW(i,j)]}$. 
Let this minimum normalized weight be referred to as $\theta_i$. % =  \min\limits_{j \in Neigh(i)}{[NW(i,j)]}$. % minimum weight edge among his or her direct 1-hop friends; This is the cut-off threshold.
\item For each $m$ of $i$'s $n$-hop friends, if $SS_n(i,m) \geq \theta_i$, the user $m$ is inserted in the candidate peer set of $i$.
Intuitively, this ensures that the social strength between $i$ and $m$, located at distance $n$ in the social graph, is at least as strong as $i$'s weakest direct tie. 
%\end{enumerate}
\squishend

We note that the algorithm expands each candidate-set using a user-specific, thus local, threshold.
Such local thresholds are needed in the distributed setting of a F2F system.

\subsubsection{ Online Presence Behavior} 

\begin{figure}
\begin{center}
      \graphicspath{{./graph/}}
      \includegraphics[scale=0.3]{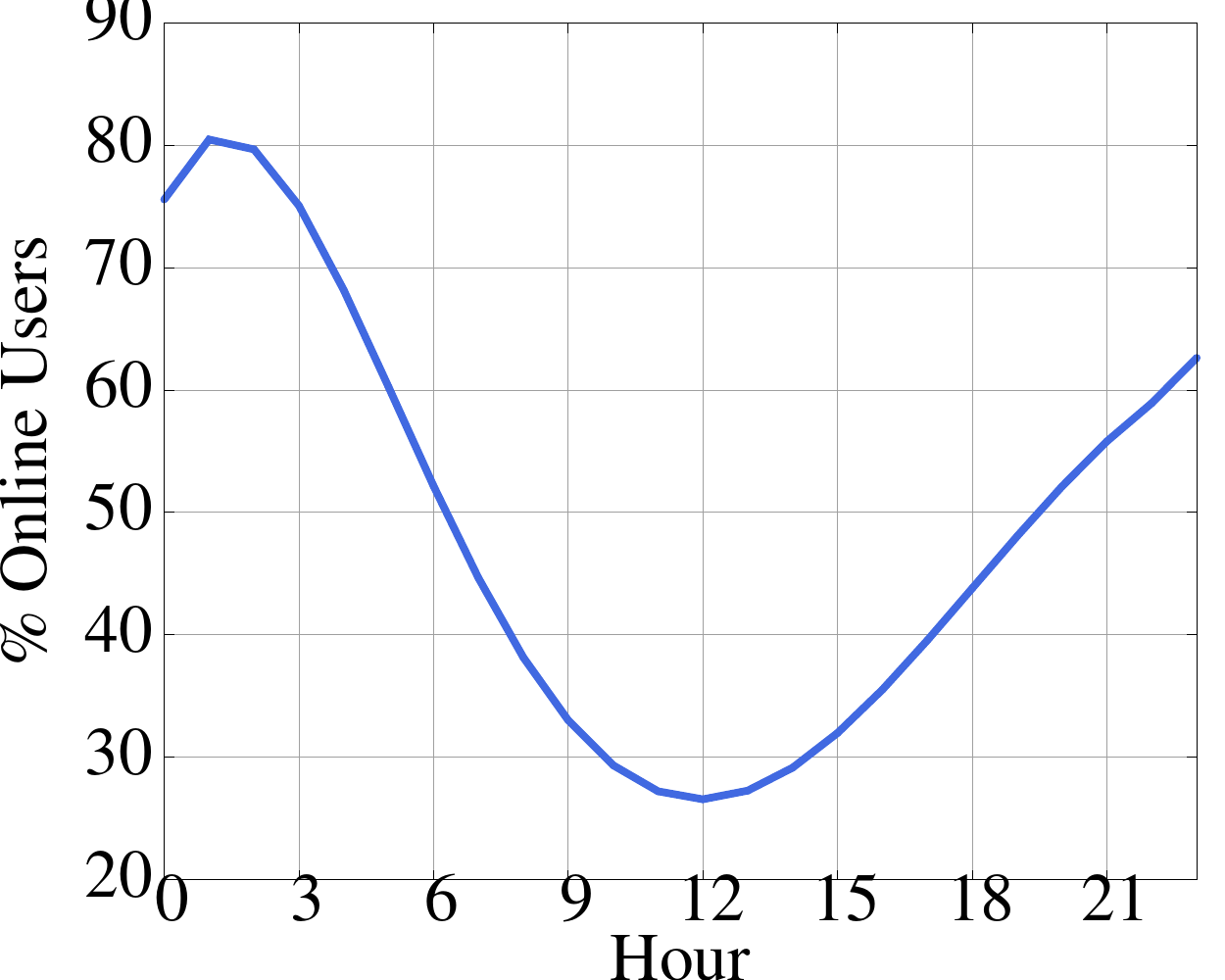}
      \caption{\label{curve} Online behavior of nodes in empirical traces of Skype.}
\end{center}
\vspace{-4mm}
\end{figure}

To estimate peer availability, we augmented each network with online presence empirically deduced from various real traces.
For CA-I and CA-II, we fit a distribution to online presence information extracted from empirical Skype traces presented in~\cite{raul2012peer}. 
The distribution was applied to each author by shifting it to match the timezone of his or her affiliation.
As seen in Figure~\ref{curve}, which plots the percentage of users online per hour of the day, at least 25\% of nodes are online at any given time, with the peak and valley occurring at about 1:00 AM and noon, respectively.

\begin{figure}[t]
\begin{center}
      \graphicspath{{./figure/}}
      \includegraphics[scale=0.55]{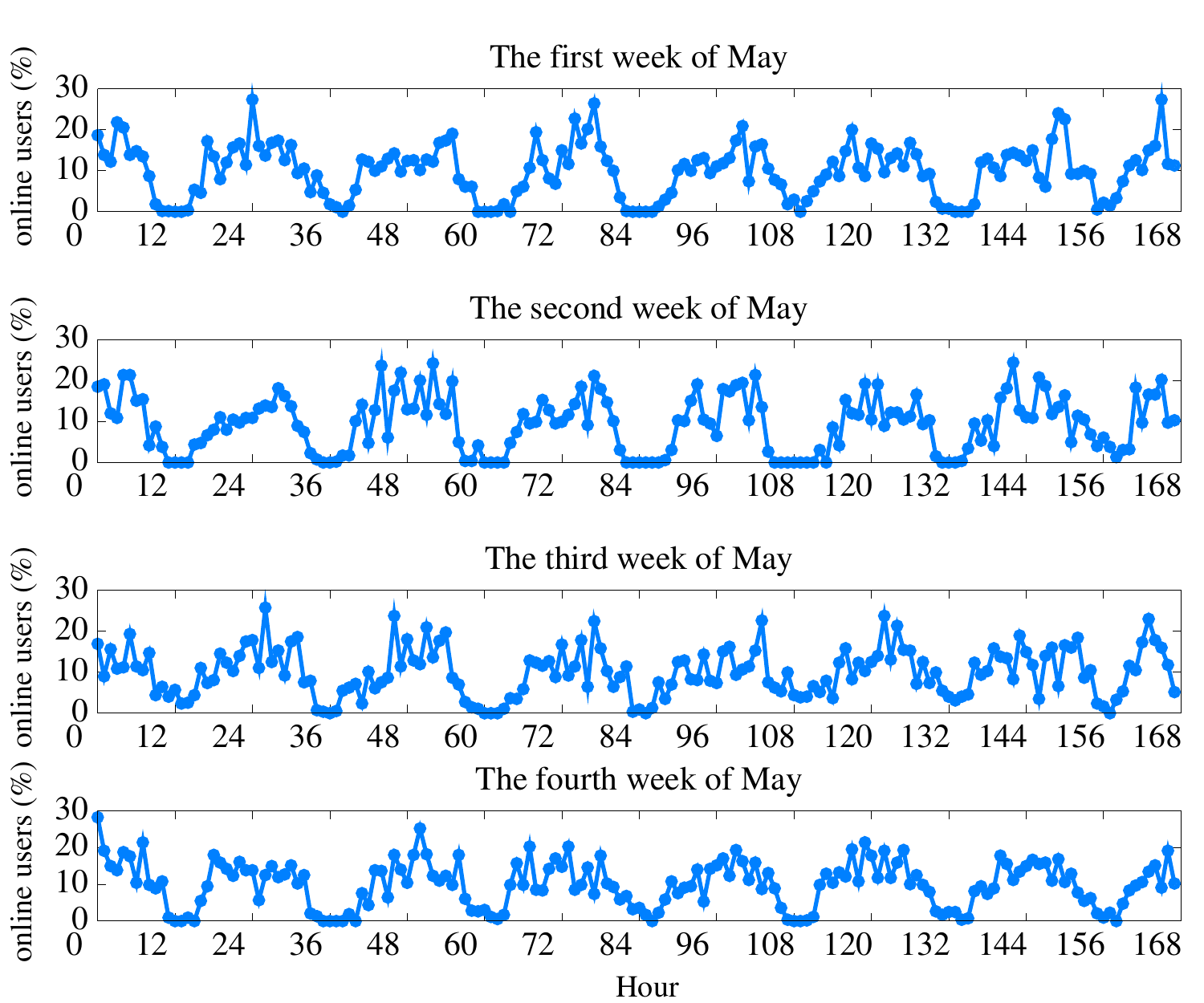}
      \caption{Online behavior of players per hour of the week in May for TF2.}
      \label{maypattern}
\end{center}
\vspace{-4mm}
\end{figure}

For the TF2 network, we used one month of the empirical playing times of the gamers.
We plot the corresponding aggregate distribution in Figure~\ref{maypattern}, which shows each week's online presence per hour for May 2011.
The distribution shows clear diurnal and domain-specific activity patterns.
As noted in~\cite{blackburn2013relationships}, gaming is not an activity conducive to multi-tasking.
Therefore, we see an elevated level of presence on weekends and during non-working hours.
Although peak presence occurs consistently in the early morning with more than 20\% of users online, there are almost no users online at noon.

%{\bf Storage Candidate Availability.} 
To determine whether the social strength selection mechanism improves the potential availability of storage resources, we measure the percentage of a node's selected candidates available throughout the day, by binning online presence into 1-hour time slots.
If a user was online at some point during a time slot, we mark him as available for that time slot.
Methods that store files in a distributed fashion such as erasure codes require $k$ storage sites to be available for retrieving a file.
Thus, we also vary the number of friends necessary for a node's storage needs to be met under such storage schemes.
We then measure the fraction of nodes who have enough candidates online to meet their needs when selected by either the 1-hop or social strength mechanisms.

\subsubsection{Data placement}

Replicating data at all friends allows a user to get maximum achievable data coverage but results in high costs for storing and transferring data to multiple copies, in particular for users with a large number of friends.
So we adapt the greedy heuristic data placement algorithm proposed in~\cite{sharma2011anempirical} to backup files with a subset of friends who can cover the maximum online time.
In this heuristic, to get maximum possible time slots coverage (e.g., 24 hours), users first pick the friend who is able to cover as many time slots as possible, then pick the second friend to maximum cover the remaining not covered time slots, and keep doing this until all the possible time slots are covered.  
%%%\ainote{Xiang, I don't understand your explanation below. Try again?}
%users pick their own critical friends (unique friends cover the only time slots) to cover all the possible time slots, if critical friends are not able to provide the maximum time slots coverage, then the user picks other online friends that covers the maximum number of the time slots not yet covered.
%\ainote{add one sentence on how this greedy heuristic works?}

%%%We utilize a day's users' online/offline status that simulated from Skype traces to implement the greedy heuristic data placement algorithm for CA-I and CA-II datasets, and use another day's status to study data availability. 
%%%Similarly, in TF2 dataset, users' uptime/downtime behavior from one week of May 2011 was used to drive the data placement algorithm, and the achieved availability in the remaining three weeks of May were studied. 
%For each week, we repeat the same experiment and aggregate all the results and average them to get an averaged data availability.

%-----------------------------
\subsection{Results}
%-----------------------------

\subsubsection{Expanding Peer Sets}
Since the most intuitive advantage of our mechanism is an increase in the number of storage candidates, we begin by evaluating \emph{how much} the candidates set is expanded.
We thus implemented $SS_n$ presented in Eq.~\ref{eq:socs} and report the size of the candidate set selected based on the expansion algorithm presented in Section~\ref{sec:expansion-alg} on the three networks described in Section~\ref{sec:datasets}.

Table~\ref{tab:expansions-stats} shows high level statistics on how candidate sets are expanded with 2- and 3-hop social distance respectively in each of our three networks.
For 2-hop expansion, 63.6\% users in CA-II and 36.6\% of players in TF2 expanded their candidate sets. 
%, expanded their candidate set, and 
Even in the sparse CA-I, 34.2\% users augmented their peer-set.
%In CA-I about 10\% of users have at least 100 candidates while in TF2 5\% of users have more than 500 candidates.
%\ainote{This doesn't make sense: I think we need to compare the same numbers with 1-hop neighborhoods: Conversely, only a small portion of users in CA-I have fewer than 100 candidates when using only direct social ties, and more than 95\% of users in TF2 have fewer than 50 candidates.}

When using only the 3-hop neighborhood to recruit peer candidates $p$ who satisfy the requirement that $SS_3(i, p) \geq \theta(i)$ %,  augmenting users' friends' candidates with further distance, 3-hop social circles, 
the expansion is still taking place in all three networks: even in the sparse network CA-I, 10.6\% users augment their friendsets and about 1\% users have expanded their candidates with more than five friends.
The denser network CA-II has more than 50\% users expanding their candidate sets, and TF2 has 27.2\% (with the number of expanded 3-hop friends being 1,032).
%%%\ainote{I don't understand this: ``with 1,032 as the maximum number of expanded peers''}. 
As expected, 3-hop augmentation is not as strong as 2 hops' since as the social distance increases, the social strength weakens.
Yet a number of users can still recruit more peers when increasing the social distance. 
Thus, using social strength for recruiting resources indirectly connected in the social graph can successfully augment users' peer-sets and potentially solve problems caused by the limited number of friends in F2F systems. %, such as reduced data availability~\cite{raul2012peer}.
%\ainote{is this the best title?}
\begin{table}[t] 
\begin{center}
\caption{Candidate set expansion via $SS_2$ and $SS_3$: percentage of expanded users, expansion nodes and rate.}\label{tab:expansions-stats}
%\begin{tabular}{|l||r|r|r|r|r|r||}
\begin{tabular}{|l||c|c|c|c|c|c|}
\hline
& \multicolumn{2}{c|}{\% expanded users} & \multicolumn{2}{c|}{expansion: med, max} & \multicolumn{2}{c|}{expansion rate: med, avg, max} \\
Net. & n=2 & n=3 & n=2 & n=3 & n=2 & n=3\\
 \hline\hline 
 CA-I & 34.2 & 10.6  & 0, 19 & 0, 19 & 0, 0.2, 1.6 & 0, 0.1, 1.1			\\ 
 \hline
 CA-II & 63.6  & 51.2  & 3, 459 & 1, 474  & 0.5, 1.3, 6.1 & 0.1, 3.1, 23.3	\\ 
\hline 
TF2 &  36.6 & 27.2  & 0, 988 & 0, 1032 & 0, 2.5, 36 & 0, 1.8, 116.5		\\
% CA-I & 34.2 & 10.6\%  & 0,19 & 0,19 & (0,0.19,1.58 & 0,0.05,1.12			\\ 
% \hline
% CA-II & 63.6  & 51.2\%  & 3,459 & 1,474  & 0.5,1.26,6.12 & 0.12,3.14,23.25	\\ 
%\hline 
%TF2 &  36.6 & 27.2  & 0,988 & 0,1032 & 0,2.49,36 & 0,1.84,116.5		\\
\hline
\end{tabular}  
\end{center}
\vspace{-4mm}
\end{table}

\ignore{
\begin{table*}[t] 
\begin{center}
\itshape  
\begin{tabular}{|l||c|c|c|c|}
\hline
Networks & Per. of expansion users & Expansion(min, median, max) & Expansion ratio (avg, median, max)\\
 \hline\hline 
 CA-I & {\itshape 10.6\%} & (0, 0, 19) & (0.05, 0, 1.12)\\ 
 \hline
 CA-II & {\itshape 51.2\%}  &( 0, 1, 474)& (3.14, 0.12, 23.25)\\ 
\hline 
TF2 &  {\itshape 27.2\%} &(0, 0, 1032) & (1.84, 0, 116.5)\\
\hline
\end{tabular}  
\end{center}
\caption{High level statistics on 3-hop candidate set expansion.}\label{tab:expansions-stats-3hop}
 \end{table*}
}

\begin{figure}[t]
\begin{center}
      \graphicspath{{./figure/}}
      \includegraphics[scale=0.25]{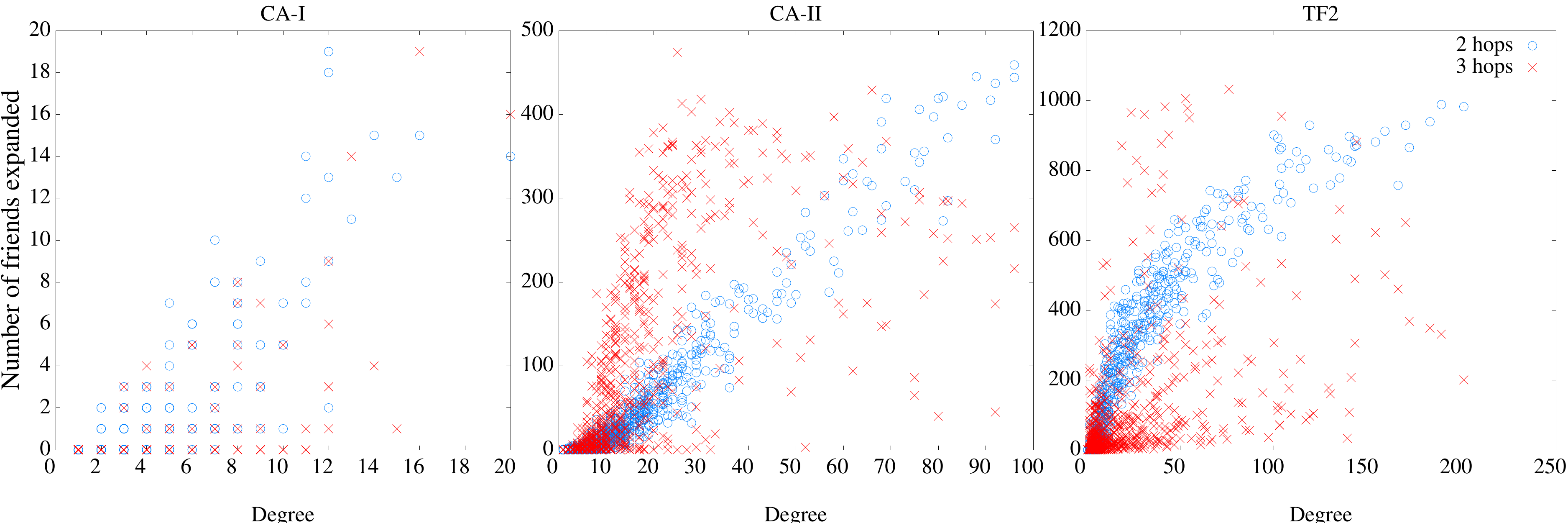}
      \caption{Expanded candidate set size as a function of node degree.}\label{fig:expansion}
\end{center}
\vspace{-4mm}
\end{figure}

Figure~\ref{fig:expansion} examines the effects of the social strength mechanism from a different perspective.
It plots the degree of a user vs. the size of her expanded candidate set.
% Users with higher degrees are able to be augmented with greater number of friends. 
For the most part, all users expand their candidate set, with CA-II showing linear growth in 2-hop distance as the user's degree increases. %\jbnote{Have you actually fit a linear function to this in case a reviewer throws a fit?}
% CA-II's friends expansion shows a linear growth as the user's degree increases.
As expected, the number of expanded friends is positively correlated to users' social degrees in both 2- and 3-hop expansion.

\subsubsection{Online Storage Availability}
 
\begin{figure}[t]
\begin{center}
      \graphicspath{{./figure/}}
      \includegraphics[scale=0.50]{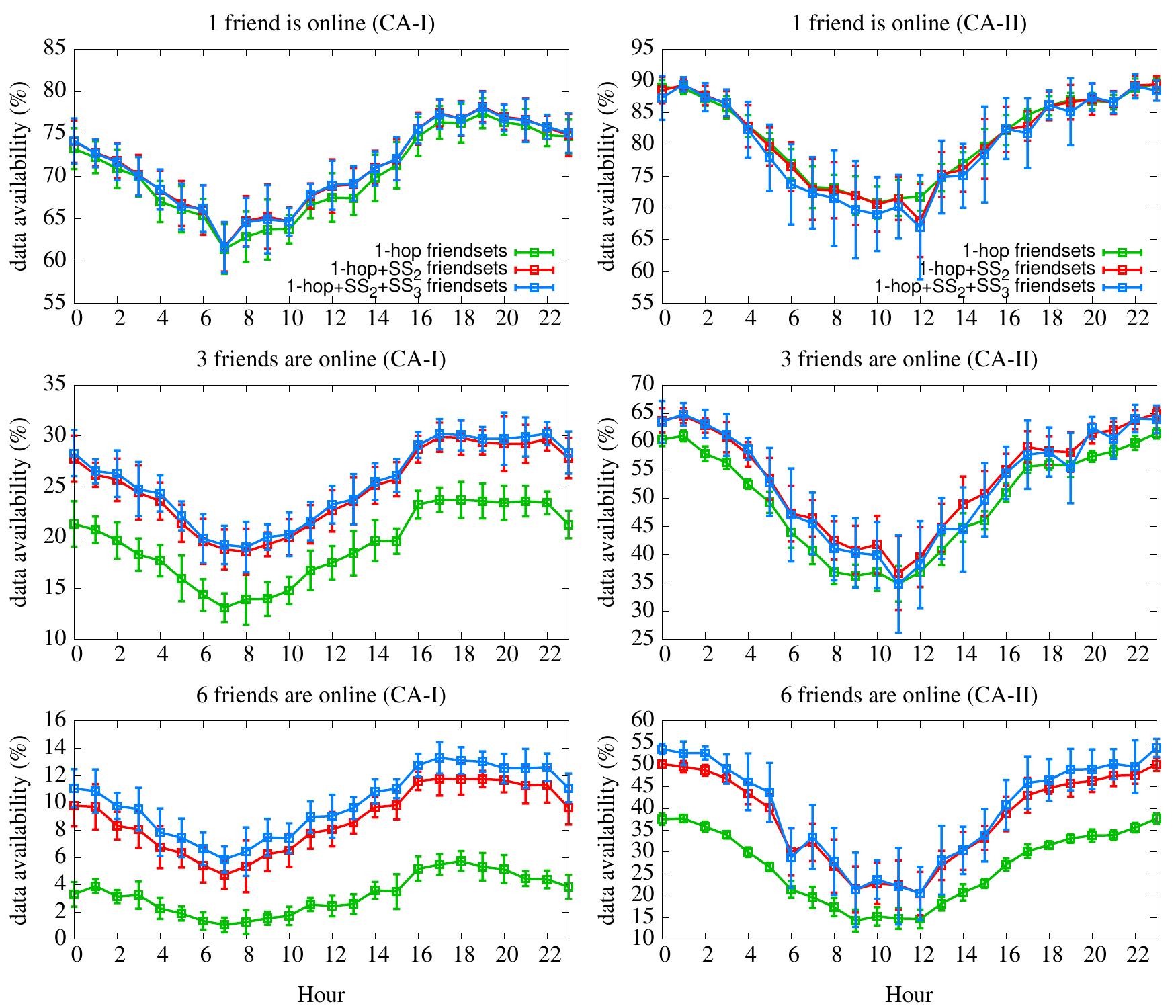}
      \caption{Average fraction of available candidates per hour for CA-I and CA-II.}
      \label{ca-all}
\end{center}
\vspace{-4mm}
\end{figure}

\begin{figure}[t]
\begin{center}
      \graphicspath{{./figure/}}
      \includegraphics[scale=0.55]{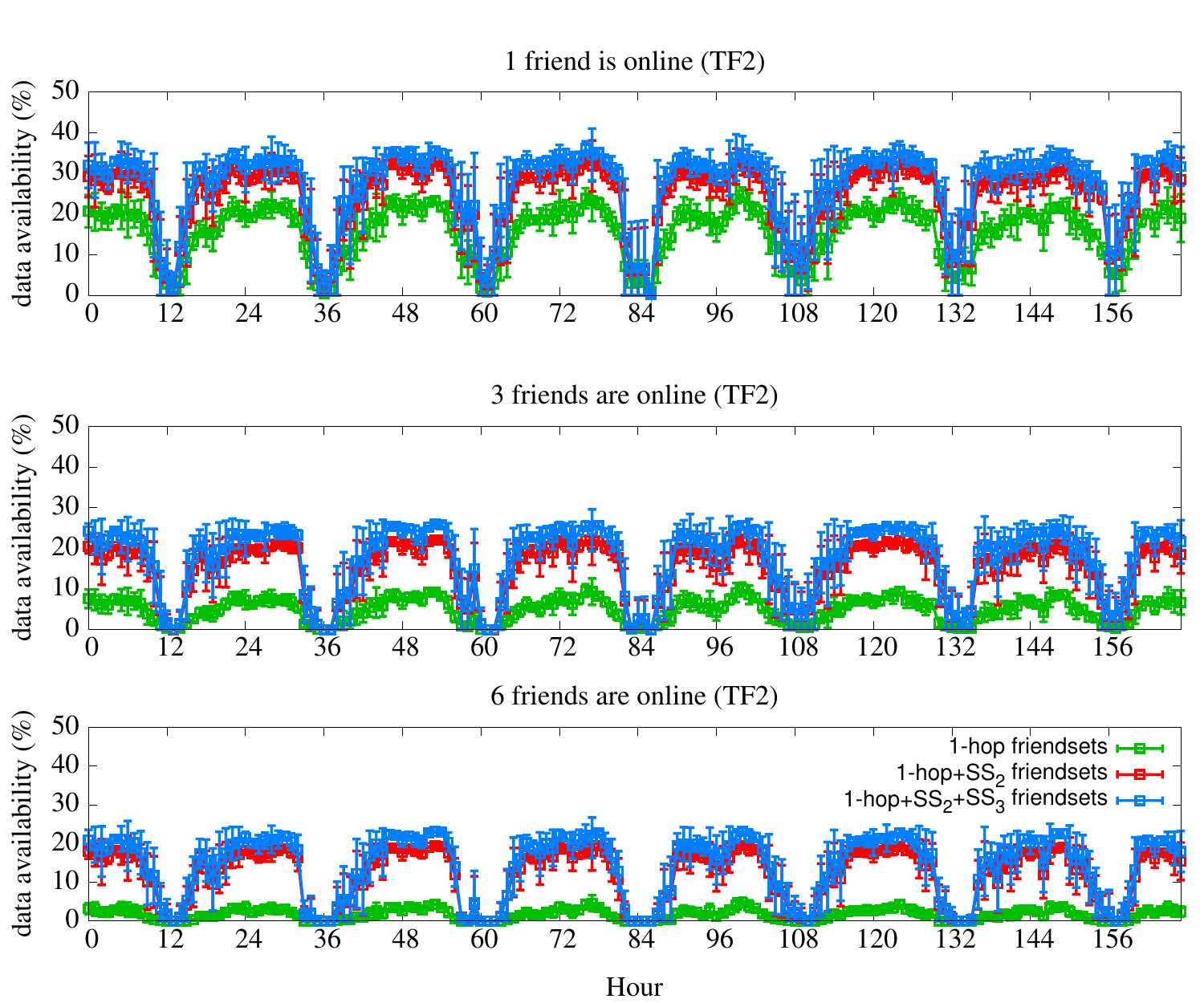}
      \caption{Average fraction of available candidates per hour of the week for TF2.}
      \label{tf2}
\end{center}
\vspace{-4mm}
\end{figure}

Figures~\ref{ca-all} and~\ref{tf2} plot the average fraction of users whose storage needs are met with the requirement that at least $k \in \{1, 3, 6\}$ candidates are online at a given time for the co-authorship networks and TF2, respectively.
% , over 24 hours based on the simulation traces in Figure~\ref{curve}. 
For CA-I and CA-II, each data point is the average of 10 times iterations.
%simulations. 
%\ainote{I don't understand this:} For TF2, every week's data are used to drive the data placement algorithm in turn and the remaining weeks are used for test studies.
%The final results are the average of all iterations with. 
Error bars represent the 95\% confidence interval.
Three scenarios are compared: storage candidates are selected only from direct social contacts; and storage candidates are selected from those located at least $n$ hops away with a strong enough social strength $SS_n$, where $n=2$ and $n=3$. 
%%%We aggregate all possible expanded peer sets for each social distance in our availability experiments, e.g., $Peerset_{\leq2}$ = $Peerset_{1 hop}$ + $Peerset_{2 hop}$ and $Peerset_{\leq3}$ = $Peerset_{1 hop}$ + $Peerset_{2 hop}$ + $Peerset_{3hop}$.

Using social strength results in a higher fraction of candidate sets meeting the storage requirements: in our experiments, social strength lead up to 20\% improvements on data availability.
In particular, when 6 friends are needed to cooperate on completing a storage task, at least 5\% higher data availability can be reached.
%%%\ainote{numbers here}.
Further, the social strength mechanism does not degrade as quickly as the 1-hop selection when increasing the number of friends that are required to be online simultaneously.
\ignore{Next, the lowest points and peak points of candidate availability in the system are consistent with the valleys and peaks of users' online behaviors.
It is reasonable that friends online patterns are correlated to their availability, and further impact the system's data availability.}
We also see that for sparse networks like CA-I, social strength over larger distance $n$ significantly improves data availability, especially when larger number of friends are required to be online simultaneously.
%when storage candidates are already covered by 2-hop expanded peer sets, 3-hop expansion does not show much positive effect on improving data availability in CA-II network but still outperform 1- and 2-hop peer sets in a sparse network such as CA-I.

Finally, CA-II shows higher levels of availability than CA-I under the same conditions.
% when given similar conditions like equal number of friends that are needed, and the same users' online and offline activities. 
This is likely because CA-II has more users with larger expanded candidate sets under the social strength mechanism than CA-I (Figure~\ref{all-distribution}).
Moreover, we note that CA-I shows better performance than TF2 under the same requirements.
In the scenario that requires at least one friend online, 73\% of users in CA-I have candidates available at midnight, compared to only 20\% of TF2 users.
One explanation could be due to the limited number of concurrent players the gaming server supports (at most 32 simultaneous players).
Another explanation is that CA-I users are spread out over multiple timezones, while most of the TF2 users are geographically close to the server to minimize latency, and thus are time synchronized in their gaming patterns. % (as the gaming application requires). 

To conclude, using datasets from co-authorship networks and a video gaming community, we show that the social strength-based mechanism more than doubles the set of storage candidates potentially motivated by social incentives, and improves the data availability of storage resources by up to 20\% with 3-hop social strength.

%Most of previous works simply classify interpersonal relations as "strong" or "weak".
%Strength is vaguely defined as a combination of the intimacy, the amount of time spent together and personal homophily.
%Forbidden triad: if person A has a strong tie to both B and C, then it is unlikely for B and C not to share a tie.
%Many research discovered that diffusion traverse greater social distance when passed through weak ties ranger than strong.
%but no one provides a method to decide who could be reached through weak ties, i.e., fine-grain calibrate weak ties.

%Diffusion of information (rumors, innovations, getting a job)--Homophily

%=============================================
\section{Using Social Strength for Predicting Information Diffusion Paths}
\label{sec:diffusion}
%=============================================

The diffusion of information is a fundamental process in OSNs.
%While it is often impossible
%\jbnote{this used to say impossible, but it doesn't make sense with rest of sentence. Please verify that it should be "possible" (or something else?).} 
%to directly observe when users propagate the information, %observing who influences whom on transmission is difficult.
Our social strength metric can be used to infer information diffusion paths between indirectly connected users in the social graph. % indirect social ties.
I.e., if we know someone who received the information at $t_0$, then we can directly predict the infected users at $t_n$ ($n \geq 2$) instead of step-wise, e.g., at $t_1$.

Predictions over such longer intervals could help OSN providers customize strategies for preventing or accelerating information spreading. 
For example, to contain rumors, OSN providers could block related messages sent to the susceptible users several time steps before the rumor arrives, or disseminate official anti-rumor messages in advance.
Similarly, advertisers could accelerate their advertisements spreading in the network by discovering who will be the next susceptible to infection.
This n-hop long paths prediction can supply more time for decision makers to contain harmful disseminations, and to choose users who are pivotal in information spreading for targeted advertisements.

%The goal of this experiment is to show that Social Strength (SS) can predict information spreading paths in OSNs.
%Notably, the predicted diffusion paths are beyond 1-hop relationships, for example 2-hop away paths, i.e., if we know a node has some information and would like to disseminate it at timestamp $t_0$, we can predict what other nodes will accept the disseminated information at timestamp $t_2$ instead of only timestamp $t_1$.

%If we could predict the diffusion paths ahead, we can watch those dangerous nodes who might accept rumors or diseases in the near future.
%In our experiments, currently, we only focus on predicting diffusion paths at 2-hop distance with single seed source.

%\subsection{Experiment Design}
This section describes our experiments of applying the social strength metric to information diffusion path prediction. % and the prediction performance.

%------------------------------
\subsection{Experimental Setup}
%------------------------------

For simplicity, we focus on predicting diffusion paths at 2- and 3-hop distance with a single seed source.
A single node is chosen as the original source of information at $t_0$.
We then predict the nodes that will accept the information at $t_n$ with the knowledge from $t_0$.
All other n-hop ($n\geq4$) paths and multiple seeds can be extended from 2 and 3 hops. 

\subsubsection{Diffusion simulation}
As ground truth, we used the widely accepted Susceptible-Infected (SI) model~\cite{newman2009networks} to simulate the diffusion process and recorded the diffusion paths, i.e., which nodes are affected during each time step.
% To obtain the information about which nodes accepted the passed on information at different timestamps (ground truths) on an OSN's information diffusion processes, we employed a widely accepted Susceptible-Infected (SI) model~\cite{newman2009networks} to simulate diffusion processes and recorded all the diffusion paths (whom are influenced or affected during each diffusion phase) and timestamps information.
%, namely, in each timestamp which nodes accepted the passed on information from its neighbors (not all neighbors but those already accepted the information in the previous phases).
The SI model is a probability-based diffusion model where nodes can be in one of two states: susceptible or infected.
We say a node has accepted the information if it is infected, and once infected, can never return to the susceptible state.
The input of the simulation is a weighted graph where weights represent a quantification of the social ties between two directly connected nodes.

For simplicity, we set the spreading probability from an infected node $v_i$ to its nearest neighbor $v_j$ proportional to the edge weight $w_{i,j}$.
Introducing a constant of proportionality $\beta$, the time-independent probability of passing information from $v_i$ to $v_j$ can be written as $P_{i,j}$ = $\beta w_{ij}$, where increasing $\beta$ results in an overall increased probability of infection.
%The most direct way is to 
We set $\beta = \frac{1}{max(w_{ij})}$ (inversely proportional to the global maximum edge weight in the graph), in which case $P_{i,j} = 1$ for the globally strongest link, and $P_{i,j} < 1$ for all others.
We set a threshold $p_0$, which is in the range between the minimum $P_{min}$ and the maximum $P_{max}$.
When $p_0 = P_{min}$, the information will disseminate to all the nodes in the graph.
Conversely, if $p_0 = P_{max}$, almost no diffusion will occur.
If $P_{ij}$ $\geq p_0$, information will spread through the edge (i, j), and not spread otherwise.
The initial seed, i.e., the node infected at $t_0$, is randomly selected at the beginning of the simulation.
% Meanwhile, there must be some seeds (information source holders) at time $t_0$ to initiate the diffusion.
% To do so, we randomly select seeds in the cold-start phase.

\subsubsection{Predicting diffusion paths via social strength}
Once we generate the ground truth from the SI model, we then use social strength to predict the path of diffusion.
We calculate social strength values between the seed and its n-hop friends, then convert the social strength values to a social rank. 
Each user has a rank list for all his or her friends according to the social strength value between said user and the friend.
%i.e., allocate a rank a node's 2-hop neighbors in terms of social strength values. 
Because of the asymmetric characteristic of the social strength metric, the converted social ranks are also asymmetric. 
\ignore{
For example, for two nodes A and B, the social rank of B from A's perspective might be 1 while the social rank of A from B is 3. 
Such asymmetric social ranks are consistent with real-world social relationships among people, which can better estimate people's social ties than symmetric social ranks.
}

After obtaining social ranks, we need a cut-off threshold to decide whether or not a node's n-hop friends will be infected at $t_n$.
The strategy we adopt here is that the social ranks from both perspectives must be high, e.g., $socialrank_n(A,B)$ and $socialrank_n(B,A)$ 
both rank among the top 10\% of both user A and B's friends.
Then, the cut-off threshold can classify a node's n-hop friends into two categories: information-accepted (infected) or information-denied nodes at $t_n$.
The intuition of this cut-off is that users will both prefer to send information to their ``closest'' social ties and will likely believe the information from their ``closest'' social ties. 

\subsubsection{Prediction evaluation} 
We compare the prediction results with the ground truth obtained from the diffusion simulation to verify the effectiveness of the social strength in predicting diffusion paths. 
We evaluate our method using three metrics.
%\begin{itemize}
\squishlist
	\item {\emph Accuracy} is the proportion of true results in the population.
	\begin{displaymath}	
		{\textstyle	
		accuracy = \frac{\#~of~true~positives+~\#~of~true~negatives}{population~size}	
		}	
		\end{displaymath}	
	\item{\emph Sensitivity}, also called the true positive rate, measures the proportion of actual positives that are correctly predicted.		
		\begin{displaymath}	
		{\textstyle	
		sensitivity = \frac{\#~of~true~positives}{\#~of~true~positives + \#~of~false~negatives}	
		}	
		\end{displaymath}		
	\item{\emph Specificity} evaluates the proportion of negatives which are correctly identified.
	\begin{displaymath}	
		{\textstyle	
		specificity = \frac{\#~of~true~negatives}{\#~of~true~negatives + \#~of~false~positives}	
		}	
		\end{displaymath}
%\end{itemize}
\squishend

%------------------------------
\subsection{Results and Evaluation}
\begin{figure}[tbhp]
\begin{center}
      \graphicspath{{./figure/}}
      \includegraphics[scale=0.6]{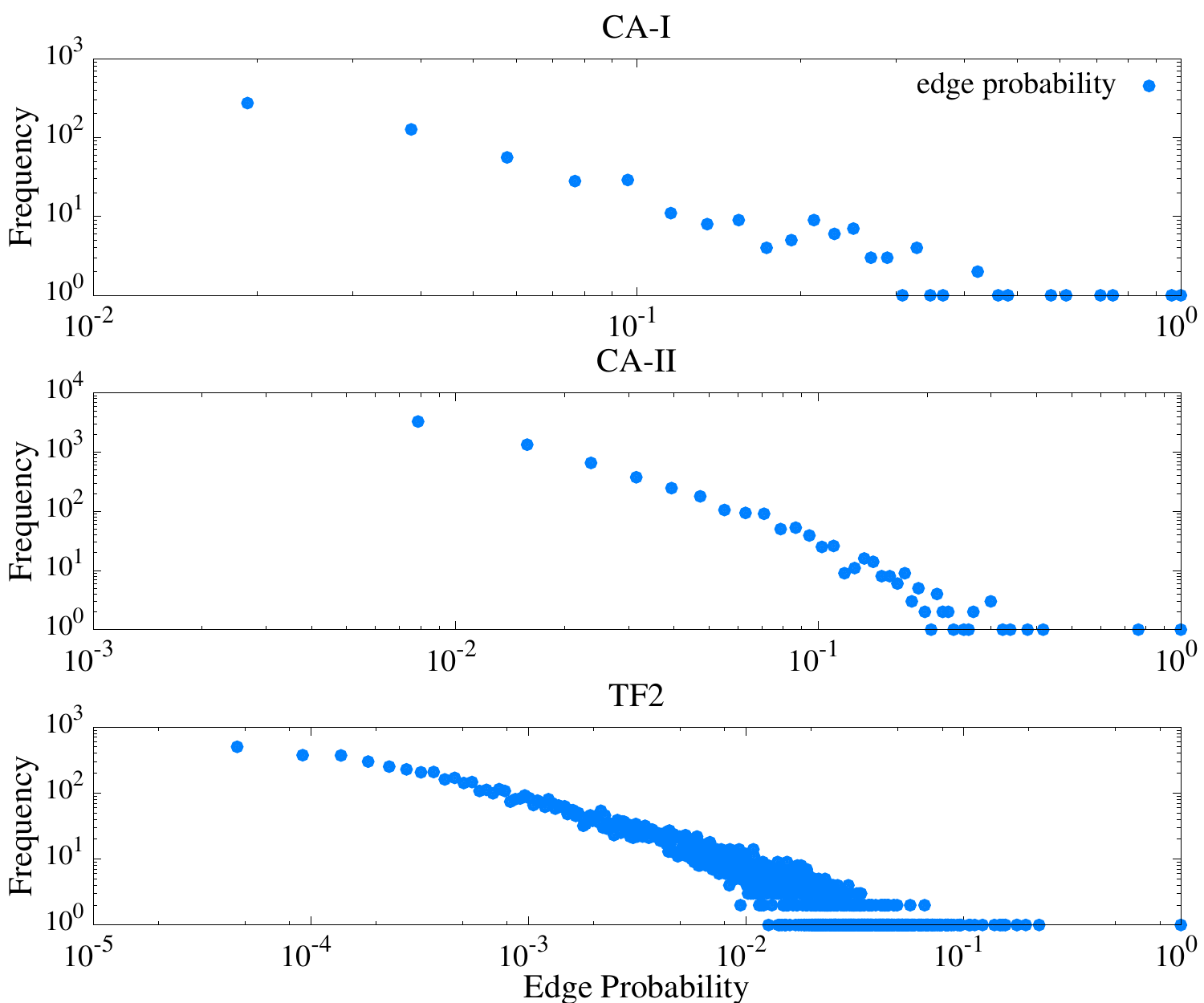}
      \caption{The frequency distribution of edge probability of CA-I, CA-II and TF2.}% in SI model.}
\label{fig:prob-dist}
\end{center}
\end{figure}

%------------------------------
We used the three datasets (CA-I, CA-II and TF2) as described in the previous section.
Results are presented as the average of 100 iterations.
We varied the threshold $p_0$ to examine the effect of prediction results in scenarios ranging from complete propagation to nearly none.
The threshold ranges we chose are based on the edge probability distribution plotted in Figure~\ref{fig:prob-dist}. For example, in CA-II, 49.3\% of edge probabilities are smaller than 0.007.
Thus, setting $p_0 <$ 0.007 results in complete diffusion in the network.
Conversely, 94.1\% edge probabilities are smaller than 0.07, so setting $p_0 >$ 0.07 leads to almost no propagation.
Therefore, we test the power of social strength for predicting diffusion paths when the threshold varied from 0.007 to 0.07.
Similarly, we choose 0.01 and 0.1 as the range of thresholds of CA-I and 0 and 0.018 for CA-II.
%(45.9\%) and 0.1(86.2\%) as the range of threshold of CA-I and 0.00004 (5.1\% edges' probabilities are smaller than 0.00004) and 0.018 (99.8\% edges' probabilities are smaller than 0.018) is the range for CA-II.
%Thus, the threshold between these boundaries represent the majority of diffusion cases.

\begin{figure}[tbhp]
\begin{center}
      \graphicspath{{./figure/}}
      \includegraphics[scale=0.5]{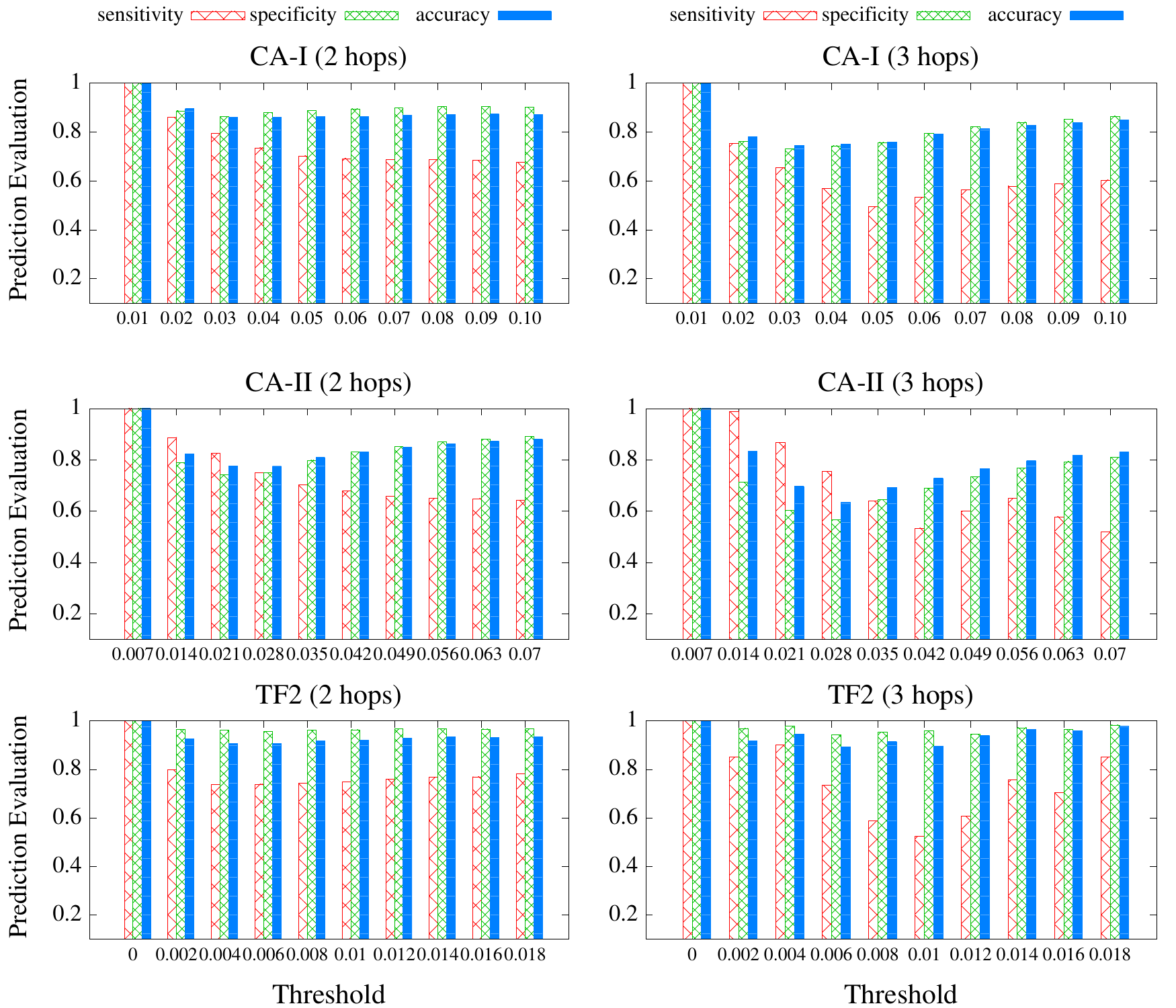}
      \caption{Sensitivity, specificity and accuracy when using social strength to predict information dissemination in CA-I, CA-II and TF2.}\label{fig:prediction-evaluation-allthree}
\end{center}
\end{figure}

The prediction results via $SS_2$ and $SS_3$ are shown in Figure~\ref{fig:prediction-evaluation-allthree}.
We see that for 2-hop social strength prediction, besides complete infection scenarios, the sensitivities are above 0.64, reaching a maximum of 0.887 in CA-II.
Also, the specificities and the accuracies in all cases are always higher than 0.74, with the highest accuracy ($0.935$) occurring in TF2 when $p_0 =$ 0.018.
Although 3-hop predictions show decreased sensitivity, specificity, and accuracy compared to 2-hop results, they remain above 0.5. 
%almost all the sensitivities are above 0.65 and the accuracies and specificities are over 0.6 when the threshold varies from 0.01 to 0.10.

\begin{figure}[tbhp]
\begin{center}
      \graphicspath{{./figure/}}
      \includegraphics[scale=0.5]{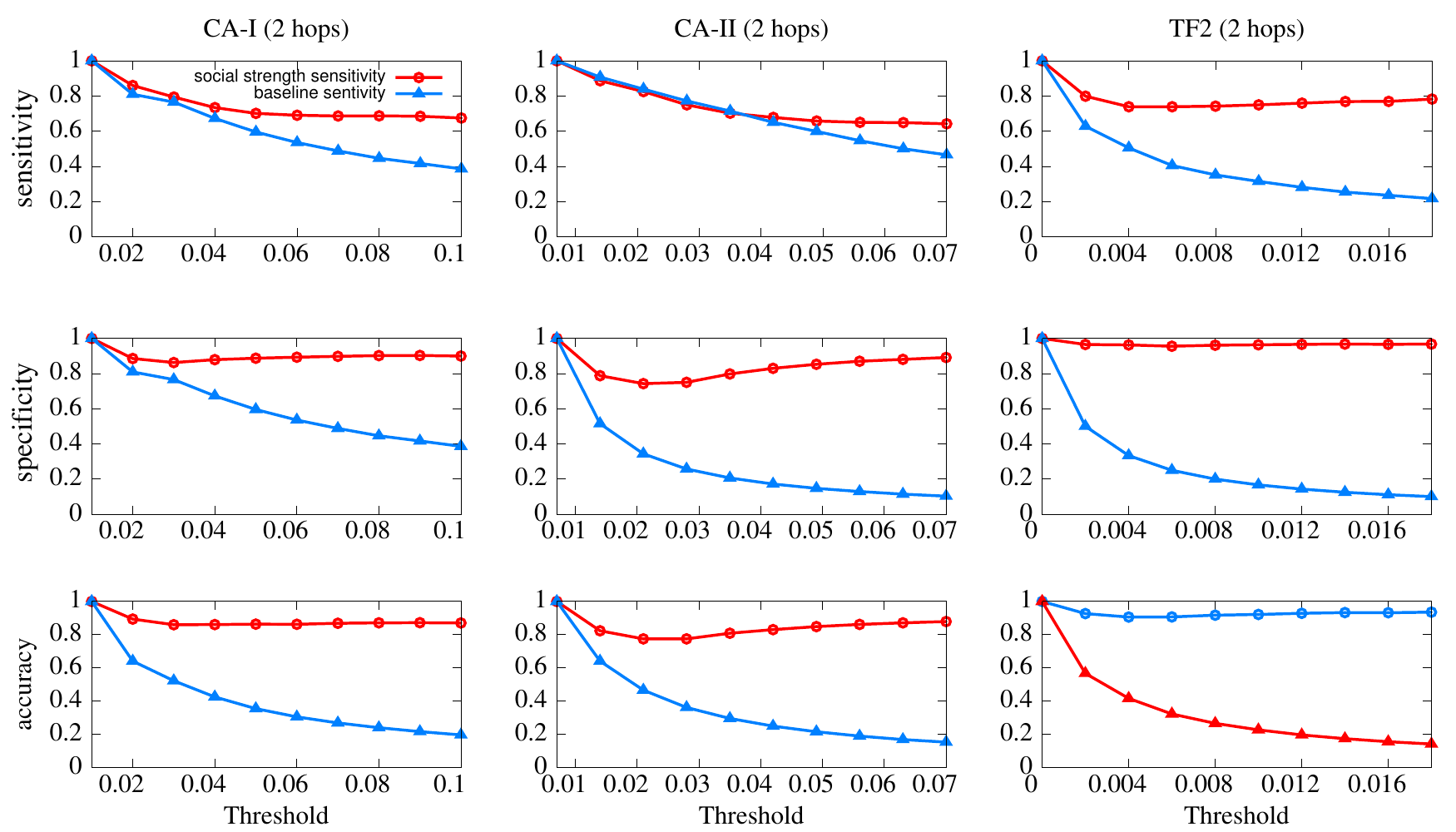}
      \caption{Comparison between the performances of prediction using 2-hop social strength ($SS_2$)) and the baseline.}\label{fig:ss-baseline}
\end{center}
\end{figure}

\begin{figure}[tbhp]
\begin{center}
      \graphicspath{{./figure/}}
      \includegraphics[scale=0.5]{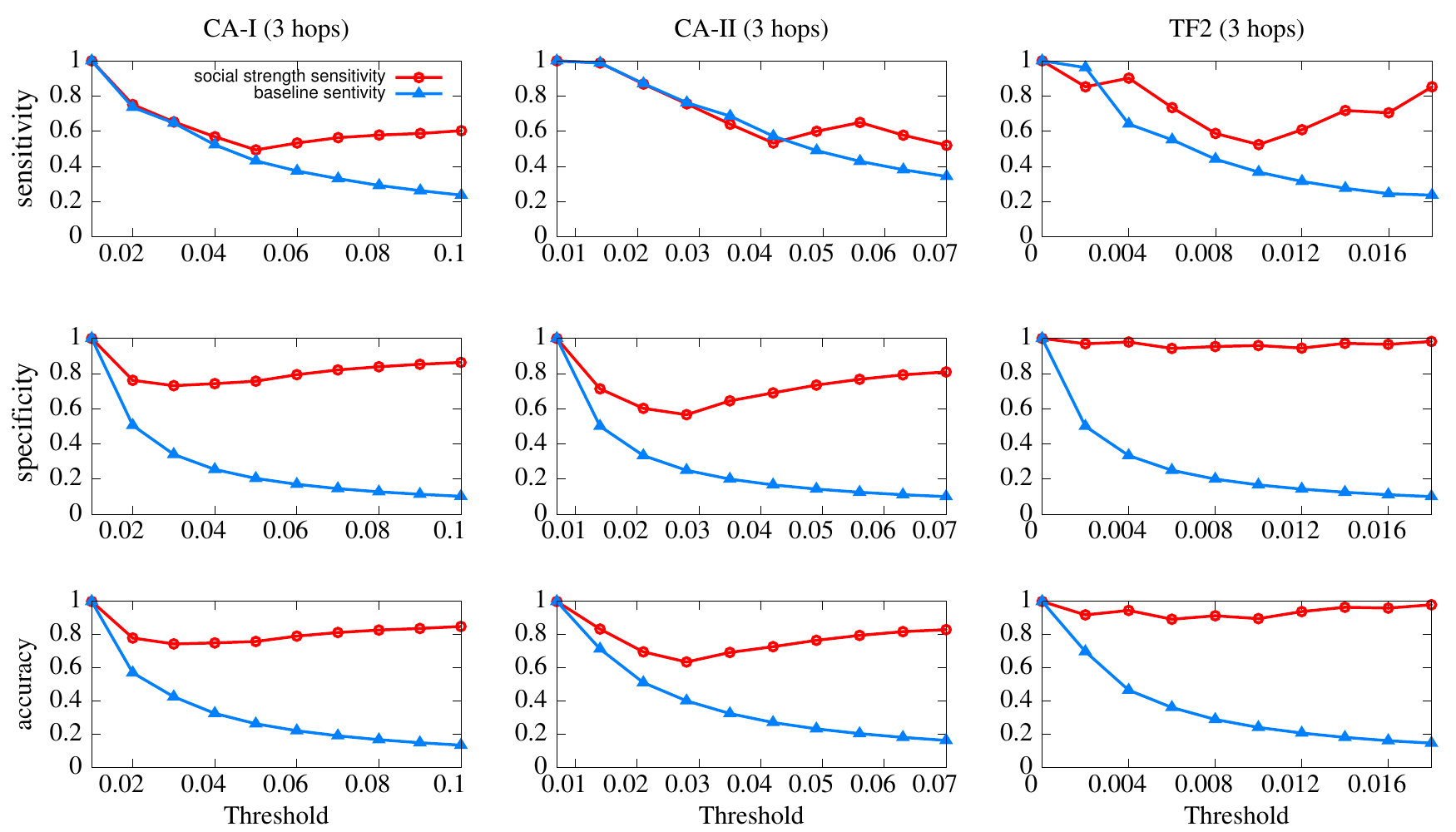}
      \caption{Comparison between the performances of prediction using 3-hop social strength ($SS_3$)) and the baseline.}\label{fig:ss-baseline-3hop}
\end{center}
\end{figure}

%{\bf Baseline method.} 
To better demonstrate social strength's effective power on inferring diffusion processes, we consider the following baseline method: assume all of a user's 2 and 3-hop friends accept the information in $t_2$ and $t_3$.
We compare our prediction results via $SS_2$ and $SS_3$ with the baseline in Figures~\ref{fig:ss-baseline} and~\ref{fig:ss-baseline-3hop}.
% The social strength where social strengths show better performance than the baseline results.

The baseline performance is very low when networks are not completely diffused while the social strength predictive method achieves peak performance at around 0.89, 0.79, and 0.82 for sensitivity, specificity, and accuracy, respectively.
It is important to note that these three networks have very different network structure (from sparse to dense), yet the performance of social strength is consistently higher than the baseline %with sensitivity, specificity, and accuracy 
in all three networks.

From all these results, we conclude that social strength is useful to predict who will be infected, or along which paths information propagates, at least 2-3 steps before a susceptible node is even in contact with an infected node.
Even if the accuracy of the prediction decreases with distance for $n=3$, there are significant benefits in being 3 time steps ahead of the infection.  

\section{Summary and Discussion}\label{sec:summary}

We introduced a social strength metric to measure the strength of indirect social ties by considering both the intensity of interactions and the number of connected paths. 
We showed that our metric is positively correlated with the Jaccard index and with the interaction frequency along direct social ties, indicating that it is an accurate quantification of the intensity of a(n indirect) social relationship.

We proved its applicability to two socially informed applications: friend-to-friend storage sharing systems and predicting information diffusion in a social graph.
Based on empirical data, our experimental evaluations demonstrate that using the social strength metric is beneficial in both cases. 
First, for the average user in the social graph, it helps identify indirectly connected peers with whom the user has a significant social strength that could act as social incentive in a resource sharing environment, thus significantly increasing the pool of resources available to the user. 
Second, because indirect ties diversify the pool of users (in this case, by covering more time zones), resource availability increases significantly.
Third, social strength accurately predicts information diffusion paths at least 2 steps ahead, which enables intervention mechanisms for rumor squelching and targeted information injection.  

A variety of socially aware applications can benefit from the social strength metric. 
%Some examples include the following: 
For example, link prediction based on social strength could discover more potentially useful contacts and improve link recommendation accuracy. 
Automatically setting default privacy controls based on social strength is likely to be more accurate than using graph distance alone. 
Employing social strength in graph partitioning will have the benefits of relying on local computation, thus allowing for more decentralized and scalable algorithms.  %can be used for allocating a node's n-hop relationships with strong social strength to the same computation machine. 
%Such social strength-based graph partitioning will decrease the delays and overheads of network communications among different machines;
%(4)~In decentralized online social networks, augmenting users' trusted friendsets with social strength not only provides more "options" for users to share files with especially in the presence of churn to mitigate the system's delays but also a more economical and privacy-guaranteed service than cloud-assisted Peer-to-Peer OSNs~\cite{mega2013cloudassisted}.
Finally, in decentralized OSNs, %, compared to cloud-assisted Peer-to-Peer OSNs~\cite{mega2013cloudassisted},
 users' augmented social strength-based friendsets could provide a more efficient and privacy-guaranteed technique to propagate updates in the presence of churn. % and mitigate the system's delays.

This work is a first step in understanding the value of and the methodology for quantifying the strength of indirect social ties. 
In addition to exploring the applicability space, there are aspects related to privacy and security that need to be understood. 
Intuitively, because of the local exploration of one's social neighborhood for computing social strength, the risks are contained, especially compared to approaches that require the global graph. 
However, a formal study of this topic is required for building a practical framework that enables the implementation and adoption of the social strength metric for indirect ties.

\ignore{
(1) Link prediction as a function of n-hop social strength;
(2) As a localized algorithm, it can be use for extending graph partitioning to n-hop social strength, i.e., allocating a node's n-hop relationships with strong social strength to the same computation machine;
(4) 
}

{
 \bibliographystyle{IEEEtran}
\bibliography{sigproc}
}

\end{document}